\pdfoutput=1
\documentclass[sigconf,table]{acmart}

\AtBeginDocument{%
  }

\copyrightyear{2024}
\acmYear{2024}
\setcopyright{rightsretained}
\acmConference[CHI '24]{Proceedings of the CHI Conference on Human Factors in Computing Systems}{May 11--16, 2024}{Honolulu, HI, USA}
\acmBooktitle{Proceedings of the CHI Conference on Human Factors in Computing Systems (CHI '24), May 11--16, 2024, Honolulu, HI, USA}
\acmDOI{10.1145/3613904.3642442}
\acmISBN{979-8-4007-0330-0/24/05}

\usepackage{xspace}
\usepackage{color}
\usepackage{listings}
\usepackage{multirow}
\usepackage{array}
\usepackage{makecell}

\newcommand{\ie}{{i.e.,}\xspace}
\newcommand{\eg}{{e.g.,}\xspace}

\newcommand{\ea}{{et~al.}\xspace}

\newcommand{\etc}{{etc.}\xspace}

\newcommand{\bpstart}[1]{\vspace{1mm} \noindent{\textbf{#1.}}}
\newcommand{\istart}[1]{\vspace{1mm} \noindent{\textit{#1:}}}
\newcommand{\ipstart}[1]{\vspace{1mm} \noindent{\textit{#1.}}}
\newcommand{\bstartnc}[1]{\vspace{1mm} \noindent{\textbf{#1}}}

\definecolor{deepblue}{cmyk}{0.9,0.75,0,0.5}
\newcommand{\code}[1]{{\color{deepblue}\texttt{#1}}}

\newcommand{\al}[1]{$\textit{#1}$}
\newcommand{\eqValue}[1]{{\color{darkgray}\textit{#1}}}
\newcommand{\eqNValue}[1]{{\color{darkgray}#1}}

\newcommand{\qcell}[2]{\makecell[l]{{\tiny \sffamily \textbf{\textcolor{darkgray}{#1}}} \\ {\scriptsize #2}}}
\newcommand{\tcell}[3]{\makecell[l]{{\tiny \sffamily \textbf{\textcolor{darkgray}{#1}}} \\ {\scriptsize #2} \\ {#3}}}

\definecolor{lightgray}{rgb}{.9,.9,.9}
\definecolor{darkgray}{cmyk}{0,0,0,0.7}
\definecolor{purple}{rgb}{0.65, 0.12, 0.82}
\definecolor{darkgreen}{cmyk}{1,0,1,0.2}

\lstdefinelanguage{JavaScript}{
  keywords={class, typeof, new, true, false, catch, function, return, null, catch, switch, var, if, in, while, do, else, case, break, let, const, await, async},
  keywordstyle=\color{darkgreen}\bfseries,
  ndkeywords={export, boolean, throw, implements, import, this},
  ndkeywordstyle=\color{darkgray}\bfseries,
  identifierstyle=\color{black},
  sensitive=false,
  comment=[l]{//},
  morecomment=[s]{/*}{*/},
  commentstyle=\color{gray}\ttfamily,
  stringstyle=\color{orange}\ttfamily,
  morestring=[b]',
  morestring=[b]"
}

\begin{document}

\title{Erie: A Declarative Grammar for Data Sonification}

\author{Hyeok Kim}
\email{hyeok@northwestern.edu}
\orcid{0000-0003-4340-4470}
\affiliation{%
  \institution{Northwestern University}
  \city{Evanston}
  \state{Illinois}
  \country{USA}
}

\author{Yea-Seul Kim}
\email{yeaseul.kim@cs.wisc.edu}
\orcid{0000-0003-1854-1537}
\affiliation{%
  \institution{University of Wisconsin-Madison}
  \city{Madison}
  \state{Wisconsin}
  \country{USA}
}

\author{Jessica Hullman}
\email{jhullman@northwestern.edu}
\orcid{0000-0001-6826-3550}
\affiliation{%
  \institution{Northwestern University}
  \city{Evanston}
  \state{Illinois}
  \country{USA}
}

\renewcommand{\shortauthors}{Kim et al.}

\begin{abstract}
Data sonification---mapping data variables to auditory variables, such as pitch or volume---is used for data accessibility, scientific exploration, and data-driven art (\eg~museum exhibitions) among others.
While a substantial amount of research has been made on effective and intuitive sonification design, software support is not commensurate, limiting researchers from fully exploring its capabilities.
We contribute \emph{Erie}, a declarative grammar for data sonification, that enables abstractly expressing auditory mappings.
\emph{Erie} supports specifying extensible \textit{tone} designs (\eg~periodic wave, sampling, frequency/amplitude modulation synthesizers), various \textit{encoding} channels, auditory legends, and \textit{composition} options like sequencing and overlaying. 
Using standard Web Audio and Web Speech APIs, we provide an \emph{Erie} compiler for web environments.
We demonstrate the expressiveness and feasibility of \emph{Erie} by replicating research prototypes presented by prior work and provide a sonification design gallery.
We discuss future steps to extend \emph{Erie} toward other audio computing environments and support interactive data sonification.
\end{abstract}




\maketitle

\section{Introduction}\label{sec:intro}

\noindent Data sonification maps data variables (\eg~height, weight) to auditory variables (\eg~pitch, loudness)~\cite{hermann2008:taxonomy,kramer1997:sonification,scaletti1994:sound}.
Sonification plays an important role in domains such as data accessibility, scientific observation, data-driven art, and museum exhibitions~\cite{supper2014:sublime}.
For people with Blindness or Vision Impairment (BVI), sonification makes it possible to access data presented on screen.
In science museums or digital news articles, data sonifications can support authoring more immersive data narratives by diversifying cues.

While sonification designs vary with their intended purposes, creating data sonification is often laborious because of limited software-wise support for auditory channels, compared to a robust set of expressive visualization toolkits (\eg~D3~\cite{bostock:d32011}, ggplot2~\cite{wickham:ggplot22010}).
An ability to express diverse designs helps creators and developers to be less constrained in making their artifacts.
Due to a lack of expressive tools for data sonification, however, many prior empirical works in accessible visualization rely on more hand-crafted methods (\eg~using Garage Band by Wang~\ea~\cite{wang2022:intuitiveness}) or solution-specific approaches (\eg~Hoque~\ea~\cite{hoque2023:naturalsound}). 
For example, Sonification Sandbox~\cite{walker2003:sandbox}'s authoring interface for data sonifications does not support expressing a sequence or overlay of multiple sonifications.
Creators of artistic sonifications or data stories need to use additional audio processing software to combine those sonifications, which requires a different set of skills.
Furthermore, those tools are not programmatically available, so it is hard to apply them to use cases with data updates or user interactions.
While several R and JavaScript libraries support creating data sonifications (\eg~DataGoBoop~\cite{datagoboop}, PlayItByR~\cite{playitbyr}, Sonifier.JS~\cite{sonifier}), they are tightly bound to the associated visualization's chart type (\eg~histogram, boxplot) or support few encoding channels (\eg~pitch only), limiting authors' potential to compose diverse data sonification designs.

To facilitate research and tool development for data sonification, we contribute \emph{Erie}, a declarative grammar for data sonification.
We developed \emph{Erie} with the goal of supporting independence from visual graphs, expressiveness, data-driven expression, compatibility with standard audio libraries, and extensibility with respect to sound design and encodings.
At high level, \emph{Erie}'s syntax specifies a sonification design in terms of \textit{tone} (the overall quality of a sound) and \textit{encoding} (mappings from data variables to auditory features).
\emph{Erie} supports various \textit{tone} definitions: 
oscillator, FM (frequency modulation), and AM (amplitude modulation) synthesizer, classical instruments, periodic waveform, and audio sampling.
Authors can specify various auditory \textit{encoding} channels, such as time, duration, pitch, loudness, stereo panning, tapping (speed and count), and modulation index.
Authors can also use \emph{Erie} to express a composition combining multiple sonifications via repetition, sequence, and overlay.
Our open-sourced \emph{Erie} player for web environments supports rendering a specified sonification on web browsers using the standard Web Audio and Speech APIs.
\emph{Erie}'s queue compiler generates an \textit{audio queue} (a scheduled list of sounds to be played), providing the potential for extending \emph{Erie} to other audio environments like C++ and R.

We demonstrate \emph{Erie}'s expressiveness by replicating accessibility and general-purpose sonification designs proposed by prior work (\eg~Audio Narrative~\cite{audioNarrative:siu2022}, Chart Reader~\cite{thompson2023:chartreader}, and news articles~\cite{vegas}). 
We provide an interactive gallery with a variety of example sonification designs.
We conclude by outlining necessary future work for \emph{Erie}, including technological hurdles, potential use cases, and blueprints for supporting \mbox{interactivity and streaming data.} 
\section{Background and Related Work}\label{sec:rw}
This work is grounded in research on data sonification and declarative grammars for data representation.

\subsection{Data Sonification}\label{sec:rw:sonification}
Data sonification or audio graph encodes data values as auditory values~\cite{hermann2008:taxonomy,kramer1997:sonification,scaletti1994:sound}. 
For example, Geiger counter maps ionizing radiation to the loudness of a sound.
Sonification is considered as one of the primary methods for data accessibility or accessible data visualization for people with Blindness and Vision Impairment (BVI).
For instance, web-based data visualization can be coupled with sonification along with alternative text descriptions. 
Yet, accessibility is not the only venue for sonification, but various fields, such as scientific data representation~\cite{andrea2005:meteorology,john1999:lifemusic,ghosh2010:particle}, data-driven art~\cite{sonificationArt}, and public engagement with science (\eg~learning~\cite{tomlinson2017solar}, museums~\cite{walker2006aquarium,dini2023:museum}), use data sonification.

\bpstart{Auditory channels}
Different auditory channels, such as pitch or volume, are physicalized into a waveform. 
We first describe a few core concepts related to a sound wave: \textit{frequency} and \textit{amplitude}.
The frequency of a sound wave refers to the number of wave cycles (\eg~a local peak to the next local peak) per second, and its unit is hertz (Hz).
A sound with a higher frequency has shorter wave cycles, and people tend to perceive it as a higher pitch.
The amplitude of a sound wave means the extent or depth of a waveform.
A larger amplitude makes a louder sound.

Commonly used channels in prior work include pitch, loudness (or volume), tapping, timing, panning, timbre, speech, and modulation index~\cite{dubus2013:review}.
\textit{Pitch} refers to how sound frequency is perceived with an ordered scale (low to high; \eg~Do-C, Re-D, Mi-E). 
\textit{Loudness} means how loud or intense a sound is, often physically measured using the unit of decibel.
\textit{Timing} is when a sound starts and stops being played; the time interval is termed \textit{duration} (or length). 
\textit{(Stereo) panning} refers to the unidimensional spatial (left to right) position of a sound by controlling the left and right volume of two-channel stereo audio.
\textit{Timbre} (or instrument, put more casually) means the quality of a sound, such as piano sound, synthesizer, bird sound, \etc
Modulation-based synthesizers (or synths), such as frequency modulation (FM) and amplitude modulation (AM), have two oscillators, a carrier for the main sound and a modulator that changes the carrier's waveform through some signal processing (simply put).
A \textit{modulation index} (MI) for such synths refers to the degree of modulation in signal processing. 
The frequencies of two oscillators generate the \textit{harmonicity} between them.

An audio mapping of a non-categorical variable can have a positive or negative \textit{polarity}.
A positive polarity scale maps a higher data value to a higher audio value (\eg~high pitch, high volume), and a negative polarity scale maps a higher data value to a lower audio value. 
While a sonification designer should be able to specify the range of an audio scale, audio scales are capped by the physical space. 
For example, the common audible frequency spectrum is known to range from 20 Hz to 20,000 Hz~\cite{audibleSpectrum}.

\bstartnc{Empirical studies in data sonification for accessibility} focus on how people with BVI interpret different auditory mappings. 
Walker~\ea~\cite{walker2010:universal,walker2002:magnitude,walker2007:consistency} extensively compared how sighted and BVI people perceive various auditory channels and the polarity of mappings for different quantitative data variables (\eg~dollars, temperature).
Recent work extends focus to other qualities of auditory mappings.
For instance, Hoque~\ea~\cite{hoque2023:naturalsound} used natural sound (\eg~bird sound) to support enhanced distinction between categorical values. 
Wang~\ea~\cite{wang2022:intuitiveness} show that BVI readers find certain audio channels to be more intuitive given visual encodings (\eg~pitch for bar heights) and given data type (\eg~quantitative, ordinal).
In their experiment, participants indicated a need for an overview of auditory scales~\cite{wang2022:intuitiveness}.
Thus, a sonification grammar should be able to express such aspects of an audio graph design definition.

\def\tabRowColorOne{e06619}
\def\tabRowColorTwo{ff9900}

\newcommand{\da}[0]{\cellcolor[HTML]{\tabRowColorOne}\color{white}{$\circ$}} 
\newcommand{\ph}[0]{\cellcolor[HTML]{\tabRowColorTwo}\color{white}{$\circ$}} 
\newcommand{\is}[0]{\cellcolor[HTML]{\tabRowColorTwo}\color{white}{$\circ$}} 
\newcommand{\isx}[0]{\cellcolor[HTML]{\tabRowColorTwo}{\tiny\color{white}{Via sampling}}} 
\newcommand{\ec}[0]{\cellcolor[HTML]{\tabRowColorOne}\color{white}{$\circ$}} 
\newcommand{\ecx}[0]{\cellcolor[HTML]{\tabRowColorOne}{\tiny\color{white}{Via preset filter}}}
\newcommand{\cl}[0]{\cellcolor[HTML]{\tabRowColorOne}\color{white}{$\circ$}} 
\newcommand{\sa}[1]{\cellcolor[HTML]{\tabRowColorOne}{\tiny\color{white}{#1}}}
\newcommand{\fl}[1]{\cellcolor[HTML]{ff9900}{\tiny\color{white}{#1}}} 
\newcommand{\rf}[0]{\cellcolor[HTML]{e06619}\color{white}{$\circ$}} 
\newcommand{\cp}[0]{\cellcolor[HTML]{ff9900}\color{white}{$\circ$}} 
\newcommand{\ap}[0]{\cellcolor[HTML]{e06619}\color{white}{$\circ$}} 

\newcommand{\tna}[1]{{\tiny #1}}

\def\tabColumnOneWidth{4.5em}
\def\tabColumnTwoWidth{3.5em}
\def\tabColumnOneToTwoWidth{8em}
\newcommand{\tabCLine}[0]{\cline{3-22}}
\newcommand{\tabCLineB}[0]{\cline{2-22}}

\begin{table*}[t]

\caption{Comparison of Erie to prior sonification toolkits. 
Abbreviations: VL (VoxLens and Sonifier.JS)~\cite{sonifier,sharif2022:sonifier},
HC (Highcharts Sonification)~\cite{walker2021:highchart},
WSS (Web Sonification Sandbox)~\cite{wss:kondak2017},
SC (Soncification Cell)~\cite{ossia:poret2023},
AAG (Apple Audio Graph)~\cite{apple:audioGraph},
DGB (DataGoBoop)~\cite{datagoboop},
PR (PlayItByR)~\cite{playitbyr},
XS (xSonify)~\cite{xsonify:candey2006},
SS (Sonification Sandbox)~\cite{walker2003:sandbox},
SY (SonifYer)~\cite{sonifyer:dombois2008},
IST (Interactive Sonification Toolkit)~\cite{istk:pauletto},
SW (Sonification Workstation)~\cite{workstation:phillips2019},
SA (SonArt)~\cite{sonart:ben2002},
L (Listen)~\cite{listen:wilson1996},
M (MUSE)~\cite{muse:lodha1997},
PS (Personify)~\cite{personify:barrass2002},
Str (Strauss)~\cite{strauss:harrison2021},
Sta (StarSound)~\cite{starsound:hannam2014},
SD (SODA)~\cite{soda:cibils2020},
Eq (Equalizer),
Pow (Power function), Sqrt (Square-root function), SymLog (Symmetric log function).}
\Description{This table has 21 columns for different sonification toolkits and 51 rows for different sonification design properties. These properties are categorized in terms of development environment, year, data, sound tone, encoding, audio filters, reference, composition, and API. Erie supports all the properties but circular pan. The other tools only support a few properties with common properties including freedom in encoding choices, pitch, loudness and stereo pan encoding channels.}

\footnotesize
\setlength{\tabcolsep}{3pt} 
\def\arraystretch{1.185}%
\resizebox{\textwidth}{!}{\begin{tabular}{|l|l|l|c|c|c|c|c|c|c|c|c|c|c|c|c|c|c|c|c|c|c|}

\hline
\multicolumn{2}{|l|}{\textbf{Category}} & \textbf{Property} &
    \textbf{Erie} & \textbf{VL} & \textbf{HC} &\textbf{WSS} & \textbf{AAG} & \textbf{DGB} & \textbf{PR} & \textbf{XS} & \textbf{SS} & \textbf{SY} & \textbf{IST} & \textbf{SW} & \textbf{SA} & \textbf{L} & \textbf{M} & \textbf{PS} & \textbf{Str} & \textbf{Sta} & \textbf{SD} \\ \hline 

\multicolumn{3}{|l|}{Environment} & 
 \tna{Web} & \tna{Web} & \tna{Web} & \tna{Web} & \tna{Swift} & \tna{R} & \tna{R} & \tna{Java} & \tna{Java} &     & \tna{Pure Data} & \tna{C++} & \tna{C++} & \tna{C++} & \tna{C} & & \tna{Python} &  &  \\ \hline

\multicolumn{3}{|l|}{Year} & 
 \tna{2023} & \tna{2022} & \tna{2021} & \tna{2017} & \tna{2021} & \tna{2020} & \tna{2011} & \tna{2006} & \tna{2004} & \tna{2008} & \tna{2004} & \tna{2019} & \tna{2002} & \tna{1996} & \tna{1997} & \tna{1995} & \tna{2021} & \tna{2020} & \tna{2014} \\ \hline

\multirow{6}{\tabColumnOneWidth}{Data} & \multirow{6}{\tabColumnTwoWidth}{Trans-form} & Aggregate & 
 \da &     &     &     &     &     &     & \da &     &     &     &     &     &     &     &     &     &     &  \\ \tabCLine
 & & Bin & 
 \da &     &     &     &     &     &     &     &     &     &     &     &     &     &     &     &     &     &  \\ \tabCLine
 & & Filter & 
 \da &     &     &     &     &     &     &     &     &     &     &     &     &     &     &     &     &     &  \\ \tabCLine
 & & Calculate &  
 \da &     &     &     &     &     &     &     &     &     &     &     &     &     &     &     &     &     &  \\ \tabCLine
 & & Density & 
 \da &     &     &     &     &     &     &     &     &     &     &     &     &     &     &     &     &     &  \\ \tabCLine
 & & Fold & 
 \da &     &     &     &     &     &     &     &     &     &     &     &     &     &     &     &     &     &  \\ \hline

\multirow{11}{\tabColumnOneWidth}{Sound tone} & \multicolumn{2}{l|}{Speech} & 
 \ph & \ph & \ph &     & \ph &     &     &     &     &     &     &     &     &     &     &     &     &     &     \\ \tabCLineB

& \multirow{8}{\tabColumnTwoWidth}{Instrument type} & Oscillator & 
 \is &     & \is & \is &     &     &     &     & \is & \is & \is & \is &     & \is &     & \is & \is &     & \\  \tabCLine
 & & FM Synth & 
 \is &     &     & \is &     &     &     &     &     & \is & \is & \is &     &     &     &     &     &     & \is  \\  \tabCLine
 & & AM Synth & 
 \is &     &     & \is &     &     &     &     &     &     &     & \is &     &     &     &     & \is &     & \is  \\  \tabCLine
 & & Musical & 
 \is &     & \is & \is &     &     &     &     & \is &     &     &     &     & \is & \is &     &     & \is & \is  \\  \tabCLine
 & & Wave & 
 \is &     &     &     &     &     &     &     &     &     & \is &     &     &     &     &     &     &     &  \\  \tabCLine
 & & Noise & 
 \is &     & \is & \is &     &     &     &     & \is &     & \is & \is &     &     &     &     & \is &     &  \\  \tabCLine
 & & Natural & 
\isx &     &     &     &     &     &     &     &     &     &     &     &     &     &     &     &     &     &  \\  \tabCLine
 & & Human vowel & 
\isx &     &     &     &     &     &     &     &     &     &     &     &     &     & \is &     &     &     & \\  \tabCLineB
 & \multicolumn{2}{l|}{Instrument sampling} & 
 \is &     &     &     &     &     &     &     &     &     &     &     &     &     &     &     & \is &     &  \\ \tabCLineB
 & \multicolumn{2}{l|}{Continuous/discrete sounds} & 
 \is &     &     &     &     &     &     &     &     &     &     & \is &     &     &     &     &     &     &  \\ \hline

\multirow{26}{\tabColumnOneWidth}{Encoding}  & \multicolumn{2}{l|}{Freedom} & 
 \ec &     & \ec & \ec &     &     & \ec & \ec & \ec & \ec & \ec & \ec & \ec & \ec & \ec &     & \ec & \ec & \ec \\ \tabCLineB
 & \multicolumn{2}{l|}{Ramping} & 
 \ec &     &     &     &     &     &     &     &     &     &     & \ec &     &     &     &     &     &     &  \\ \tabCLineB

& \multirow{19}{\tabColumnTwoWidth}{Channels} & Time-start & 
 \ec & \ec &     &     & \ec & \ec & \ec & \ec &     &     &     &     & \ec &     & \ec & \ec & \ec & \ec & \\  \tabCLine
& & Time-end & 
 \ec &     &     &     &     &     &     &     &     &     &     &     &     &     &     &     &     &     &  \\  \tabCLine
& & Duration & 
 \ec &     &     &     &     &     &     &     &     &     & \ec &     &     & \ec &     &     &     &     & \ec  \\  \tabCLine
& & Speed/Tempo & 
 \ec &     &     &     &     &     & \ec & \ec &     &     &     &     &     &     & \ec &     &     &     &  \\  \tabCLine
& & Count & 
 \ec &     &     &     &     &     &     &     &     &     &     &     &     &     &     &     &     &     &  \\  \tabCLine
& & Pitch/detune & 
 \ec & \ec & \ec & \ec & \ec & \ec & \ec & \ec & \ec &     & \ec &     & \ec & \ec & \ec & \ec & \ec & \ec & \ec \\  \tabCLine
& & Loudness & 
 \ec &     & \ec & \ec &     &     &     & \ec & \ec &     & \ec & \ec & \ec & \ec & \ec & \ec & \ec &     & \ec \\  \tabCLine
& & Stereo pan & 
 \ec &     & \ec & \ec &     & \ec &     &     & \ec &     & \ec & \ec &     & \ec &     &     &     &     &  \\  \tabCLine
& & Circular pan & 
     &     &     &     &     &     &     &     &     &     &     &     &     &     &     &     & \ec &     &  \\  \tabCLine
& & Timbre & 
 \ec &     &     &     &     &     &     &     &     &     &     &     &     &     & \ec & \ec &     &     & \\  \tabCLine
& & MI & 
 \ec &     &     &     &     &     &     &     &     &     &     &     &     &     &     &     &     &     &  \\  \tabCLine
& & Harmonicity & 
 \ec &     &     &     &     &     &     &     &     &     &     &     &     &     & \ec & \ec &     &     & \\  \tabCLine
& & Reverb & 
 \ec &     &     &     &     &     &     &     &     &     &     &     &     &     &     &     &     &     &  \\  \tabCLine
& & Speech & 
 \ec &     &     &     &     &     &     &     &     &     &     &     &     &     &     &     &     &     &  \\  \tabCLine
& & Highpass filter & 
\ecx &    & \ec &     &     &     &     &     &     &     &     &     &     &     &     &     &     &     &  \\  \tabCLine
& & Lowpass filter & 
\ecx &    & \ec &     &     &     &     &     &     &     &     &     &     &     &     &     &     &     &  \\  \tabCLine
& & Envelope & 
\ecx & \ec & \ec &     &     &     &     &     &     &     &     & \ec &     &     &     &     &     &     &  \\  \tabCLine
& & Distortion & 
\ecx &    & \ec &     &     &     &     &     &     &     &     & \ec & \ec &     &     &     &     &     &  \\  \tabCLine
& & Repeat & 
\ec  &    &     &     &     &     &     &     &     &     &     &     &     &     &     &     &     &     &  \\  \tabCLineB
& \multicolumn{2}{l|}{Custom channels} & 
\ec  &    &     &     &     &     &     &     &     &     &     &     &     &     &     &     &     &     &  \\ \tabCLineB

& \multirow{4}{\tabColumnTwoWidth}{Scale} & Data type & 
 \cl &     &     &     &     &     &     &     &     &     &     &     &     &     &     & \cl &     &     & \cl \\ \tabCLine
& & Domain/range & 
 \cl &     & \cl &     & \cl &     & \cl & \cl & \cl &     &     &     & \cl & \cl &     &     &     &     & \cl  \\ \tabCLine
& & Polarity & 
 \cl &     & \cl &     &     &     &     & \cl & \cl &     &     & \cl &     &     &     &     &     &     &  \\ \tabCLine
& & Transform & 
 \sa{Log/Pow/Sqrt/SymLog} & 
           &     &     &     &     &     & \sa{Log} &
                                                     &     &     &     &     &     &     &     &     &     &  \\ \hline

\multicolumn{3}{|l|}{\multirow{4}{\tabColumnOneToTwoWidth}{Audio filters}} & 
\fl{Biquad} & 
           & \fl{High/lowpass} &
                       &     &     &     &     &     & \fl{Eq}
                                                           &     & \fl{Eq}    
                                                                       &     &     &     &     &     &     & \fl{Eq}  \\
\multicolumn{3}{|l|}{} & 
 \fl{Compressor} &    
           &     &     &     &     &     &     &     &     &     &     &     &     &     &     &     &     &  \\ 
\multicolumn{3}{|l|}{} & 
 \fl{Envelope} &
           &     &     &     &     &     &     &     &     &     &     &     &     &     &     &     &     &  \\ 
\multicolumn{3}{|l|}{} & 
 \fl{Distorter} &  
           &     &     &     &     &     &     &     &     &     &     &     &     &     &     &     &     &  \\ \hline

\multicolumn{2}{|l|}{\multirow{2}{\tabColumnOneWidth}{Reference}} & Tick & 
 \rf &     & \rf & \rf &     & \rf &     &     &     &     &     &     &     &     &     & \rf &     & \rf & \\ \tabCLine
\multicolumn{2}{|l|}{} & Audio legend & 
 \rf &     &     &     &     &     &     &     &     &     &     &     &     &     &     &     &     &     &  \\ \hline

\multicolumn{2}{|l|}{\multirow{3}{\tabColumnOneToTwoWidth}{Composition}} & Sequence & 
 \cp &     &     &     &     &     &     &     &     &     &     &     &     &     &     &     &     &     &  \\ \tabCLine
 \multicolumn{2}{|l|}{} & Overlay & 
 \cp &     &     &     &     &     &     &     &     &     &     & \cp &     &     &     &     &     &     & \cp  \\ \tabCLine
 \multicolumn{2}{|l|}{} & Repetition & 
 \cp &     &     &     &     &     &     &     &     &     &     &     &     &     &     &     &     &     &  \\ \hline

\multicolumn{3}{|l|}{API} & 
 \ap & \ap &     &     & \ap & \ap & \ap &     &     &     &     &     &     &     &     &     & \ap &     &  \\ \hline
\end{tabular}}
\label{tab:comparison}
\end{table*}

\subsection{Sonification Tools and Toolkits}\label{sec:rw:toolkits}
Prior work has proposed \textbf{sonification tools} for accessibility support for data visualizations.
For example, iSonic~\cite{sonification:zhao2008}, a geospatial data analytic tool, offers various audio feedback for browsing maps, such as using stereo panning to provide a spatial sense of the geospatial data point that a user is browsing.
iGraph-Lite~\cite{iGraphLite:ferres2013} provides keyboard interaction for reading line charts, and Chart Reader~\cite{thompson2023:chartreader} extends this approach to other position-based charts and supports author-specified ``data insights'' that highlight certain parts of a given visualization and read out text-based insight descriptions.
Siu~\ea~\cite{audioNarrative:siu2022} propose an automated method for splitting a line chart into several sequences and adding a template-based alternative text to each sequence.
Agarwal~\ea~\cite{agarwal:sonify} provide a touch-based interaction method for browsing data sonifications on mobile phones. 
While prior sonification research has focused on use of non-speech sound, accessibility studies underscore combining speech and non-speech sound to design audio charts.

Beyond supporting accessibility, others proposed \textbf{sonification toolkits} created for developers or creators to directly make data sonifications.
This prior tooling motivates a design space for sonification toolkits, such as the distinction between instrument and audio channels, needs for programming interfaces, and the utility of audio filters.
However, existing tools often provide compartmentalized support for creating expressive and data-driven sonifications as summarized in \autoref{tab:comparison}.
For example, sonification designs supported by DataGoBoop~\cite{datagoboop} and PlayItByR~\cite{playitbyr} are strongly tied to underlying chart type (\eg~histogram, box plot), limiting the freedom in choosing auditory encoding channels.
Sonifier.js~\cite{sonifier,sharif2022:sonifier} offers limited audio channels, time and pitch.
Sonification Sandbox~\cite{walker2003:sandbox} and its successors~\cite{walker2021:highchart,wss:kondak2017} support more encoding channels, but developers need to use external sound editors to sequence or overlay multiple sonifications that they created using the interface, requiring a different stack of skills. 
Furthermore, many existing tools lack application programming interface (API) support, making it difficult for users to personalize or customize sonification designs with their preferred encoding channels or instruments.
To achieve greater expressiveness with APIs, developers could use audio programming libraries, such as Tone.js~\cite{tonejs}, but they have to manually scale data values to auditory values, which can be a substantial hurdle for those with limited audio skills.
These tools also lack support for scale references (\eg~tick, scale description), making it harder to decode audio graphs they generate.

Our work provides a declarative grammar for data sonification, \emph{Erie}, as a programmatic toolkit and abstraction that developers can use to express a wide range of sonification designs. 
\emph{Erie} supports various common encoding channels (time, duration, pitch, loudness, tapping, panning, reverb, and speech), verbal descriptions, tone sampling, and composition methods (repeat, sequence, and overlay), making it a good basis for use in the development of future sonification software. 

\subsection{Declarative Grammar}\label{sec:rw:grammar}
Declarative programming is a programming paradigm where a programmer provides an abstract specification (or spec) describing the intended outcome and a compiler executes to generate the outcome. 
In this paradigm, declarative grammar defines rules for how to write a program. 
Many data-related fields, such as data visualization and statistics, have widely adopted declarative grammars. 
In data visualization, Wilkinson~\cite{wilkinson:2012grammar} proposed the \textit{grammar of graphics} as a systematic way to describe a visualization design specification.
Based on the \textit{grammar of graphics}, ggplot2~\cite{wickham:ggplot22010} for R and D3.js~\cite{bostock:d32011} and Vega stacks (Vega~\cite{satyanarayan:vega2016}, Vega-Lite~\cite{satyanarayan:vega-lite2017}) for JavaScript are widely used declarative grammars for creating general-purpose data visualizations.

Declarative grammars add value by providing internal representations and compilers for user applications, particularly when directly manipulating the targeted physical space is challenging like audio environments for sonification~\cite{joyner2022:challenge}.
For example, some sonification toolkits (\eg~\cite{istk:pauletto}) adopt visual programming languages to allow for visually and interactively authoring data sonification, and those visual programming languages are backed by some kind of declarative expressions.
For example, Quick and Hudak~\cite{donya:haskell2013} provide graph-based expressions that allow for specifying constraints to automatically generate music.
Implementing a sonification from scratch requires a sophisticated skill set for controlling electronic audio spaces (\eg~connecting audio filters, timing sounds, \etc).
To facilitate sonification software development, our work contributes a declarative grammar for data sonification, \emph{Erie}, and compilers for web environments built on standard audio APIs.
\section{Gaps in Sonification Development Practices}\label{sec:practice}

\noindent To motivate our design of \emph{Erie} with awareness of existing practices used in developing data sonification, we surveyed recently published data sonification tutorials and designs. 
To understand practices being shared among sonification developers, we collected nine online tutorials for coding sonifications by searching with keywords like ``sonification tutorial,'' ``audio graph tutorial,'' or ``sonification code.''
To see techniques beyond tutorials, we inspected 24 data sonifications with code or detailed methodology descriptions from Data Sonification Archive\footnote{\url{https://sonification.design/}} that were published from 2021 through 2023.
This collection included tutorials and designs created by active sonification contributors like Systems Sound\footnote{\url{https://www.system-sounds.com/}} and Loud Numbers\footnote{\url{https://www.loudnumbers.net/}}.
We include the list of the sonification tutorials and designs we collected in Supplementary Material.
We tagged sonification tutorials and designs in terms of software or libraries used, functionality of code written by the creators (\eg~scale functions, audio environment settings), and output formats (\eg~replicability of designs, file formats).
Overall, this preliminary survey identified that \textbf{developers currently rely on ad-hoc approaches due to the lack of expressive sonification approaches}.

\bpstart{Converting to auditory values then connecting to music software}
Most tutorials (7 out of 9) introduced \textit{music programming libraries} like music21, PyGame, Tone.js, sequenceR, Max, Sonic Pi, and MIDIFile, and most (15 out of 24) sonification designs used them. 
These libraries take as input auditory values like pitch notes or frequencies, volumes, and time durations.
That is, developers still need to define scale functions that convert data values to auditory values, requiring an understanding of physical properties of different auditory variables. 
For example, the ``Sonification 101'' tutorial\footnote{\url{https://medium.com/@astromattrusso/sonification-101-how-to-convert-data-into-music-with-python-71a6dd67751c}} describes how to map data points to notes with a four-step procedure.
First, a developer normalizes the data point into a range from 0 to 1, then multiples by a scalar to keep them in a certain range.
Third, the developer specifies a list of notes to map data points to.
Last, they write a for loop to convert each data point to the corresponding note from the list.
On the other hand, a tutorial by Propolis\footnote{https://propolis.io/articles/making-animated-dataviz-sonification.html} introduces a linear scale function.

Then, developers need to connect those computed values to other music libraries by configuring custom instruments.
To be able to create custom instruments using low-level libraries like MIDITime or Tone.js, the developer needs to have professional skills like how to import and control audio samples and what audio nodes to control to adjust different audio properties.
For instance, common sonification encodings like gain, pitch, and distortion level are governed by different audio nodes.
More experienced professional creators chose to use more advanced music software like Ableton Live, Supercollider, and Touch Designer that enable live performances or art installations.

\bpstart{Difficulty in reusing sonification designs}
Whether created programmatically or not, many existing sonification cases are shared as multimedia files (audios or videos).
This practice makes it harder to inspect how they were created in terms of data-to-music scales, instrument details, \etc.
Even if a sonification's codes are available, it is often hard to reuse the custom code because developers have to manually inspect the code in terms of different variable names to locate where to make changes for their designs.  
For example, to change the domain, range, and transformation type (\eg~sqrt, log) of a certain scale, then they have to find the relevant lines and manually change them by writing something like a linear scale function (\eg~\code{aScaleFunction(x) \{return min(1600, max((log(x)-log(30))/(log(500)-log(30))*1600, 200);\}}), \\ which is not always straightforward, particularly for less experienced sonification developers. 
This difficulty in reusing custom code is also widely known among visualization practitioners~\cite{battle2022:d3}.

\section{Design Considerations}\label{sec:consideration}
Leveraging prior empirical studies, sonification toolkits (\autoref{tab:comparison}), and development practices (\autoref{sec:practice}), we developed the \emph{Erie} grammar and compiler as a toolkit for sonification developers with the following considerations in mind.

\bpstart{(C1) Be independent}
Many existing sonification libraries that provide APIs are strongly tied to visual forms, such that they support sonifying a particular visualization instead of authoring a sonification.
While this approach can make it easy to generate sounds, it prevents sonification creators from exploring the many alternative designs one might generate by directly expressing audio graphs. 
Furthermore, it ignores different tasks implied by similar visualization designs.
For example, point marks can be a scatterplot for assessing correlation or a residual plot for judging model fit, potentially calling for different sonification designs.
We designed the \emph{Erie} grammar to be independent of visual forms to maximize design possibilities. 

\bpstart{(C2) Be expressive}
To support independently creating various sonification designs, it must be possible to express different sound qualities, auditory channels, and combinations of multiple sonifications.
Expressive toolkits enable researchers and developers to navigate a variety of design ideas.
Thus, \emph{Erie} supports specifying different sound designs (\eg~instrument types, discrete vs. continuous sounds) and different auditory channels for data encoding, and also allows for specifying sonification sequences and overlays. 

\bpstart{(C3) Be data-driven}
Sonification can be a useful tool for enhancing presentations of data in other modalities (\eg~visualization), in addition to standing on its own. 
Creating sonification often starts with implementing ad-hoc functions to convert data to audio properties as shown earlier.
Under the assumption that \emph{Erie}'s users may have limited understanding and skill with respect to acoustic engineering and audio programming, it makes more sense to be able to declare data-to-audio conversions with a few configuration terms.
Consequently, we designed \emph{Erie}'s syntax to express \textit{data} instead of \textit{sound} by leveraging the \textit{grammar of graphics}~\cite{wilkinson:2012grammar} and its popular implementations~\cite{wickham:ggplot22010,satyanarayan:vega2016,satyanarayan:vega-lite2017}, such as their scale expressions for encoding channels.

\bpstart{(C4) Be extensible}
A toolkit may not be able to support all potential cases in advance, particularly when the design space is unlimited. 
\emph{Erie} allows for sampling audio files, configuring FM and AM synths, and defining periodic waves (combining multiple sine and cosine waves). 
Furthermore, \emph{Erie} provides a method to define and connect custom audio filters (\eg~distortion, biquad filters) that can have extra auditory encoding channels.

\bpstart{(C5) Be compatible with standards}
The expressiveness and extensibility criteria are constrained by specific audio environments.
As different display media affect the resolution of images, sound representations are highly sensitive to audio environments, such as processing capacities and equipment.
Thus, compatibility with the standards of a targeted environment is critical, similar to how we use SVG or Canvas for web visualizations.
We consider two standards for sonification: (1) physical units and (2) rendering standards.
First, \emph{Erie}'s queue compiler generates a scheduled list of sound items using standard auditory units (\eg~Hz and musical notes for pitch, the panning range from $-1$ to $1$) so as to be used in other audio environments (\eg~external music software).
Our \emph{Erie} player for web employs the Web Audio and Speech APIs to enable cross-browser experience.
\section{\emph{Erie} Grammar}\label{sec:grammar}

\begin{figure*}
    \centering
    \includegraphics[width=\textwidth]{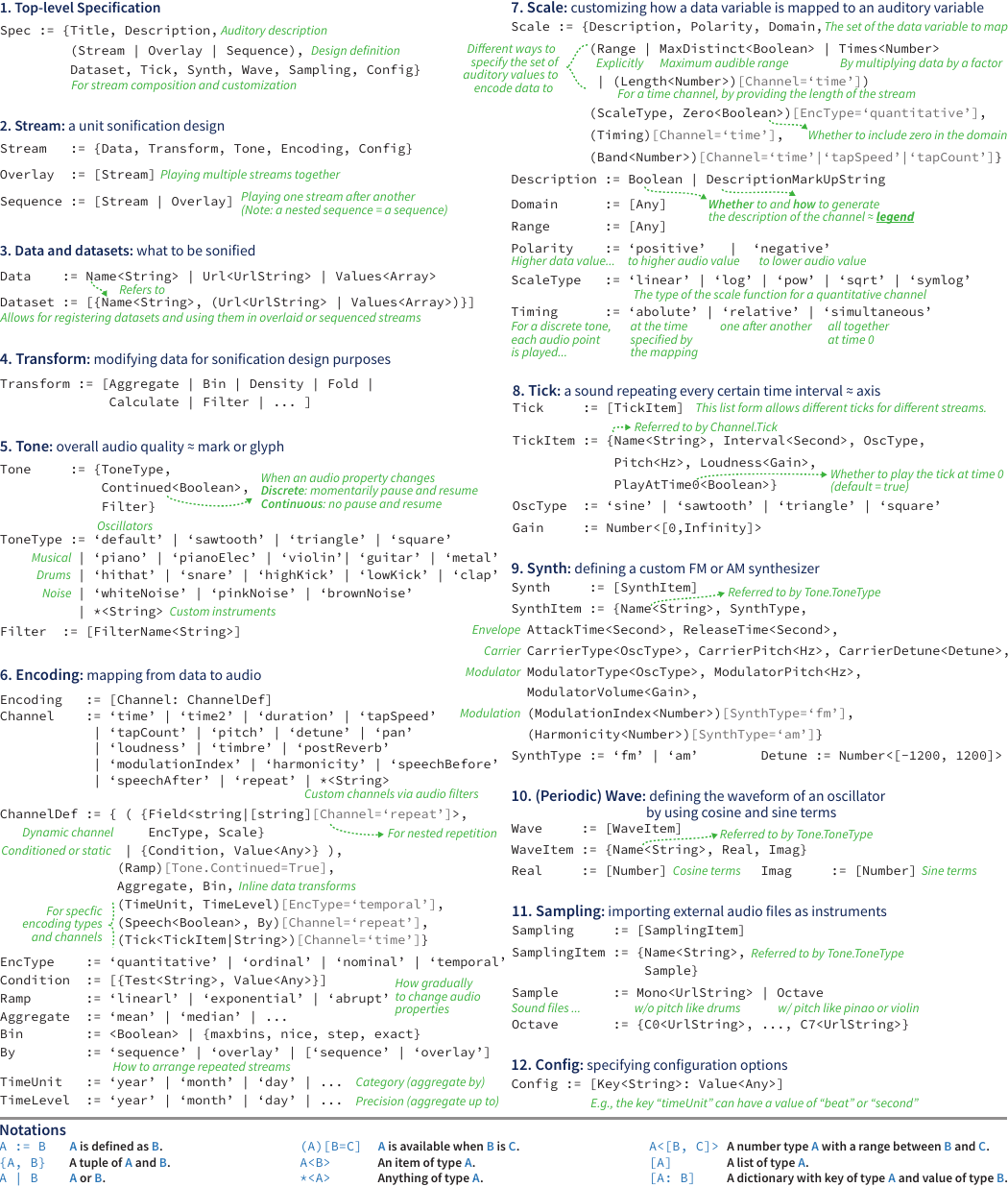}
    \caption{The formal definition of Erie. For applicable elements, roughly analogous visualization elements are denoted by $\approx$~signs.}
    \Description{This figure describes the formal specification of Erie in twelve parts. They are 1. Top-level specification. 2. Stream. 3. Data and datasets. 4. Transform. 5. Tone. 6. Encoding. 7. Scale. 8. Tick. 9. Synth. 10. Periodic wave. 11. Sampling. And, 12. Config.}
    \label{fig:grammar:formal}
\end{figure*}

\noindent We formally describe the syntax of the \emph{Erie} grammar to show how \emph{Erie} is designed to be \textbf{expressive (C2)} and \textbf{data-driven (C3)}. 
At a high level, \emph{Erie} expresses a sonification design using a sound instrument (\al{tone}) and mappings from data to auditory values (\al{encoding} channels).
After walking through an example case, we describe how \emph{Erie} expresses a data sonification design, including top-level specification, stream, data input and transform, tone, encoding, stream composition, and configuration.
The formal definition of \emph{Erie} is provided in \autoref{fig:grammar:formal}.
In describing \emph{Erie}, we distinguish \textit{developer}s who create sonifications from \textit{listener}s who listen to sonifications. 
For details, refer to the Appendix and the documentation\footnote{\url{https://see-mike-out.github.io/erie-documentation/}}.

\subsection{A Walkthrough Example}\label{sec:grammar:walkthrough}
To help imagine how \emph{Erie} works in specifying a sonification design, we introduce a simple auditory histogram for a quantitative data variable, \al{miles per gallon} with a range from five to 50, from a \al{`cars.json'} dataset~\cite{vegadataset}.
In this sonification, \textit{miles per gallon} is discretized into nine bins by five miles, and the bins are communicated by mapping them to time.
The count (aggregation) of each bin is mapped to pitch.

To construct this example using \emph{Erie}, we first specify the data to sonify by providing its URL:
\begin{align*}
\textit{data} = \{ \textit{url} = \eqValue{cars.json} \}
\end{align*}
Then, we need data \al{transform}s for binning and count aggregation.
The below expression creates bins for the \al{miles per gallon} \al{field} using default binning options (\al{auto}).
This operation defines two additional fields for the start and end point of each bin.
The expression further assigns \al{miles-per-gallon-bin} to the name of bucket start points (\al{as}) and \al{miles-per-gallon-bin-end} to the name of end points (\al{end}). 

\begin{align*}
\textit{bin}=\{ & \textit{field}=\eqValue{miles-per-gallon}, 
 \textit{auto}=\eqValue{true},  \\
 & \textit{as}=\eqValue{miles-per-gallon-bin}, \\
 & \textit{end}=\eqValue{miles-per-gallon-bin-end} \}
\end{align*}
For the count aggregation, the below expression specifies doing a \al{count} \al{op}eration, and names the resulting field \al{count}.
To count the values for each bucket, this expression sets a \al{group-by} field to the bin start point field (\al{miles-per-gallon-bin}) generated by the previous bin transform.
\begin{align*}
\textit{aggregate}=\{ & \textit{op}=\eqValue{count}, \textit{as}=\eqValue{count}, \\
& \textit{group-by}=\eqValue{miles-per-gallon-bin} \}
\end{align*}
To have the results of the bin transform feed-forward to the count aggregation, these two transforms are ordered as:
\begin{align*}
\textit{transform} = [\textit{bin}, \textit{aggregate}]
\end{align*}
Applying these transforms to the \al{`cars.json'} data results in \autoref{tab:grammar:walkthrough:tab}.

\begin{table}[h]
\caption{The results of data transforms in \autoref{sec:grammar:walkthrough}.}\label{tab:grammar:walkthrough:tab}
\Description{This table shows the results of data transformations for binning as described in Section 5.1. There are three columns for miles per gallon bin, miles per gallon bin end, and count. There are nine data rows for nine bins.}
\def\arraystretch{1.15}%
\arrayrulecolor{lightgray}
    \centering
    \small
    \begin{tabular}{lll}
    \hline
        \textbf{\al{miles-per-gallon-bin}} & \textbf{\al{miles-per-gallon-bin-end}} & \textbf{\al{count}} \\ \hline
        5 & 10 & 1 \\ \hline
        10 & 15 & 52 \\ \hline
        15 & 20 & 98 \\ \hline
        20 & 25 & 78 \\ \hline
        25 & 30 & 77 \\ \hline
        30 & 35 & 56 \\ \hline
        35 & 40 & 27 \\ \hline
        40 & 45 & 8 \\ \hline
        45 & 50 & 1 \\ \hline
    \end{tabular}
\end{table}

Second, we need to define how to sonify the specified data in terms of overall qualities (\al{tone}) and auditory mappings (\al{encoding}).
We indicate that the sound should be segmented or discrete:
\begin{align*}
\textit{tone} = \{ \textit{continued} = \eqValue{false} \}
\end{align*}
Then, we need three encoding channels: when to start each sound (\al{time}), when to end it (\al{time2}), and its \al{pitch}.
The \al{time} channel encodes the bin start points (\al{miles-per-gallon-bin}):
\begin{align*}
\textit{time} = \{ & \textit{field}=\eqValue{miles-per-gallon-bin}, 
 \textit{type}=\eqValue{quantitative}, \\
& \textit{scale}=\{\textit{length}=\eqNValue{4.5}\}
\}
\end{align*}
The above expression also specifies that the \al{time} channel encodes a \al{quantitative} variable and that the total \al{length} of the auditory histogram is \al{4.5} seconds. 
We want to finish each bin's sound with respect to the bucket's endpoint.
Because bins' start and end points are in the same unit and scale, we use an auxiliary \al{time2} channel:
\begin{align*}
\textit{time2} = \{ & \textit{field} = \eqValue{miles-per-gallon-bin-end}, \textit{type}=\eqValue{quantitative}
\}
\end{align*}
Note that this \al{time2} channel has no \al{scale} expression because it uses the same scale as the \al{time} channel.
Next, we encode the \al{count} of each bin to a \al{pitch} channel in a way that a higher count is mapped to a higher pitch (\al{positive} \al{polarity}), using the below expression:
\begin{align*}
\textit{pitch} = \{ & \textit{field} = \eqValue{count},  \textit{type}=\eqValue{quantitative}, \\
& \textit{scale}=\{ 
\textit{domain}=\eqNValue{[0, 100]},
\textit{range}=\eqValue{[220, 660]}, \\
& \qquad \qquad
\textit{polarity}=\eqValue{positive} \}
\}
\end{align*}
This expression further specifies that this \al{pitch} channel maps a \al{domain} (from 0 to 100) to a pitch frequency \al{range} (from 220Hz--A4 note to 660Hz--A6 note). 
These three encoding channels are combined as:
\begin{align*}
\textit{encoding} = \{ & \textit{time}, \textit{time2}, \textit{pitch} \}
\end{align*}

Lastly, the above expressions are combined into a \al{spec} as:
\begin{align*}
\textit{spec} = \{ & \textit{data}, \textit{transform}, \textit{tone}, \textit{encoding}  \}
\end{align*}
This \al{spec} results in the sonification output shown in \autoref{tab:grammar:walkthrough} (see Supplementary Material for the actual audio).
The equally-sized bins are mapped to the start and end times, and the aggregated counts by each bin is mapped to the pitch frequencies. 

\begin{table}[h]
\caption{The sonification output for an auditory histogram in \autoref{sec:grammar:walkthrough}. 
``\#'' indicates the playing order of each part.
Units: seconds (start, end, duration) and Hz (pitch).
``Sine'' means a sinusoidal oscillator.}\label{tab:grammar:walkthrough}
\Description{This table describes the audio queue of the resulting auditory histogram as specified in Section 5.1. The first queue item is a speech queue for "start playing." The second queue item is a tone queue for the nine histogram bins with their auditory properties including start time, end time, duration, timbre, and pitch in Hertz. The last queue is a speech queue for "finished."}
\footnotesize
\def\arraystretch{1.15}%
\arrayrulecolor{gray} 
\rowcolors{2}{white}{black!10!}
\noindent\begin{tabular}{|p{1.25em} p{3em} p{15.7em}|}
\hline
\textbf{\#} & \textbf{Type} & \textbf{Sound} \\ \hline
1 & Speech & Start playing. \\ \hline
2 & Tone & 
\makecell[l]{
\begin{tabular}{@{}lllll@{}}
\qcell{Start}{0} & \qcell{End}{0.5} & \qcell{Duration}{0.5} & \qcell{Timbre}{Sine} & \qcell{Pitch}{224.4}  \\
\hline
\qcell{Start}{0.5} & \qcell{End}{1} & \qcell{Duration}{0.5} & \qcell{Timbre}{Sine} & \qcell{Pitch}{448.8}  \\
\hline
\qcell{Start}{1} & \qcell{End}{1.5} & \qcell{Duration}{0.5} & \qcell{Timbre}{Sine} & \qcell{Pitch}{652.2}  \\
\hline
\qcell{Start}{1.5} & \qcell{End}{2} & \qcell{Duration}{0.5} & \qcell{Timbre}{Sine} & \qcell{Pitch}{563.2}  \\
\hline
\qcell{Start}{2} & \qcell{End}{2.5} & \qcell{Duration}{0.5} & \qcell{Timbre}{Sine} & \qcell{Pitch}{558.8}  \\
\hline
\qcell{Start}{2.5} & \qcell{End}{3} & \qcell{Duration}{0.5} & \qcell{Timbre}{Sine} & \qcell{Pitch}{466.4}  \\
\hline
\qcell{Start}{3} & \qcell{End}{3.5} & \qcell{Duration}{0.5} & \qcell{Timbre}{Sine} & \qcell{Pitch}{338.8}  \\
\hline
\qcell{Start}{3.5} & \qcell{End}{4} & \qcell{Duration}{0.5} & \qcell{Timbre}{Sine} & \qcell{Pitch}{255.2}  \\
\hline
\qcell{Start}{4} & \qcell{End}{4.5} & \qcell{Duration}{0.5} & \qcell{Timbre}{Sine} & \qcell{Pitch}{224.4}  \\
\end{tabular} 
}\\ \hline
3 & Speech & Finished. \\ \hline
\end{tabular}
\end{table}

\subsection{Top-Level Specification and Stream}\label{sec:grammar:stream}
We first define a simple, single data sonification specification in \emph{Erie} (a \al{spec}, hereafter) as a tuple of \al{stream}, \al{dataset}, \al{tick}, \al{synth}, \al{wave}, \al{sampling}, \al{title}, \al{description}, and \al{config}: 
\begin{align*}
\textit{spec} := 
\{ & \textit{stream}, \textit{dataset}, \textit{tick}, \textit{synth}, \textit{wave}, \textit{sampling}, \\
 & \textit{title}, \textit{description}, \textit{config} \}
\end{align*}
The curly brackets $\{$ $\}$ indicate a tuple of elements.

A \al{stream} represents a unit sonification design, consisting of \al{data} (what to sonify), \al{transform} (operations to the data), \al{tone} (overall sound quality), and \al{encoding} (mappings from data to sound values):
\begin{equation*}
\textit{stream} := \{ \textit{data}, \textit{transform}, \textit{tone}, \textit{encoding} \}
\end{equation*}
To pre-define and reuse elements in multiple \al{stream}, a developer can use different lists of named objects for \al{dataset}, \al{tick}, \al{synth} (synthesizers), \al{wave} (periodic wave), and \al{sampling} (using external audio files as a \al{tone}).
A developer can specify a speech-based \al{title} and \al{description} that are played before the audio graph.
The \al{config} of a \al{spec} configures a sonification design, such as the speed of speech (\al{speech rate}) and whether to skip playing the \al{title} text (\al{skip title}). 

\subsection{Data, dataset, and Transform}\label{sec:grammar:data}
A sonification \al{stream} must have data to sonify and \emph{Erie} supports three methods to do so: providing the \al{URL} of a data file, providing an array of \al{values}, or providing the \al{name} of a predefined dataset in the \al{dataset} object.
\begin{equation*}
\textit{data} := \textit{URL}~|~\textit{values}~|~\textit{name},
\end{equation*}
where the vertical bar sign $|$ denotes `or'. 
A \al{dataset} object consists of the named definitions of data items using \al{URL} or \al{values}.
\begin{equation*}
\textit{dataset} := [ \{ \textit{name}, \textit{URL} \} ~|~ \{ \textit{name}, \textit{values} \} ]
\end{equation*}
The square brackets $[$ $]$ denote a list of elements.

After pre-processing the data to sonify, a developer may need to perform additional, simple data transforms for sonification design purposes, such as the binning for the auditory histogram in the walkthrough. 
The developer can list transform definitions in a \al{transform} object.
In the walkthrough, for example, the \al{bin} transform created new data variables for the start and end points of each bin, and the \al{count} \al{aggregate} reshaped the data with a new variable for the count of each bin.

\subsection{Tone}\label{sec:grammar:tone}
To set the baseline sound of a sonification stream, a developer needs to specify the sound \textit{tone}.
A tone is roughly analogous to a mark or glyph in a visualization given that data values are mapped to its properties like pitch.
\emph{Erie} expresses the \al{tone} of a stream using an instrument \al{type} (\eg~piano, FM or AM synth), an indicator of whether a sound is \al{continued}, and a set of audio \al{filter}s. 
\begin{equation*}
\textit{tone} := \{ \textit{type}, \textit{continued}, \textit{filter} \}
\end{equation*}
An instrument \al{type} can be specified by its name, such as `sawtooth' (oscillator) or `violin', where default is a sinusoidal oscillator in our implementation.
If a sound is \al{continued}, two sound points are connected without a pause.
For more diverse audio expressions, the developer can provide audio \al{filter}s like distortion or equalizer.

\subsection{Encoding}\label{sec:grammar:encoding}
The \al{encoding} of a stream defines how data variables are mapped to different auditory properties (\eg~pitch and loudness) of a \al{tone}.
\emph{Erie} supports three classes of channels: dynamic, conditioned, and static. 
A \al{d}ynamic \al{channel} encodes a data variable (or field) to the respective auditory property.
It is defined in terms of a data \al{field} from the \al{stream}'s data, the data type of an encoding (\al{enc-type}), its \al{scale} details, its \al{ramp}ing method, and inline data transform options (\al{aggregate} and \al{bin}):
\begin{align*}
\textit{channel}_d := \{ \textit{field}, \textit{enc-type}, \textit{scale}, \textit{ramp}, \textit{aggregate}, \textit{bin} \}
\end{align*}
The data type of encoding (\al{enc-type}) can be either \al{quantitative}, \al{ordinal}, \al{nominal}, or \al{temporal}, reflecting common data types.
For a continuous tone, a \al{ramp}ing method specifies how to smoothly transition one auditory value to another. 
The transition can be abrupt (no-ramping), linear, and exponential.

A developer may need to emphasize certain data values by making them sound different instead of encoding every data value using a scale.
In the walkthrough, suppose that the developer wants to indicate bins with more than 80 counts using a louder sound.
Supporting such cases, a \al{c}onditioned \al{channel} has a \al{condition} list for special values and a \al{value} for the others.
\begin{align*}
\textit{channel}_c := \{ \textit{condition}, \textit{value}, \textit{ramp} \}
\end{align*}
The \al{condition} is a list of \al{test} conditions and desired \al{value}s.
\begin{align*}
\textit{condition} := [ \{ \textit{test}, \textit{value} \} ],
\end{align*}
where if a data value meets a \al{test} condition, then the specified \al{value} is assigned. 
Then, the above example can be expressed as:
\begin{align*}
\textit{loudness} = \{ & \textit{value}=\eqNValue{0.5},  \\
 & \textit{condition}=[ \{ \textit{test}=\eqNValue{(\textit{datum}.\textit{count}>80)}, 
\textit{value}=\eqNValue{1} \} ] \}
\end{align*}
Lastly, a \al{s}tatic channel only needs a \al{value} (\ie~$\textit{channel}_s := \{ \textit{value} \}$).

\subsubsection{Scale}
The \al{scale} of a dynamic encoding channel essentially consists of the \al{domain} (data values to map) and \al{range} (audio values to be mapped) of the mapping.
From the walkthrough, the domain of $[0, 100]$ and the range of $[220, 660]$ of the pitch channel compose a linear function $f(x) = (660-220) \times \frac{x}{100} + 220$.
There are shortcuts for defining a \al{range}.
When \al{max-distinct} is set to \al{true}, then the widest possible range is used (\eg~the lowest to highest human-audible pitch).
The \al{times} multiplies each data value by itself to compute auditory values. 
To verbally describe the scale, a developer can provide \al{description} using a markup expression (see \autoref{sec:appendix:reference}), analogous to a legend in a visualization.
A baseline \al{scale} is formally defined as:
\begin{align*}
\textit{scale} := \{ \textit{domain}, ( \textit{range}~|~\textit{max-distinct}~|~\textit{times}), \textit{description} \}
\end{align*}
For a quantitative variable, the developer can further specify \al{scale-type} (\eg~square-root, log, and exponential), the inclusion of \al{zero} point, and \al{polarity}:
\begin{align*}
\textit{scale}_q := \{ \ldots, \textit{polarity}, \textit{scale-type}, \textit{zero} \}
\end{align*}
An ellipsis ($\ldots$) denotes the baseline properties.

\subsection{Composition}
Combining multiple \al{stream}s is necessary to create rich auditory data narratives (\eg~\cite{thompson2023:chartreader,audioNarrative:siu2022}).
For example, a stream for vote share can be repeated to provide statistics for different regions. 
Alternatively, two streams, one for vote shares and one for the number of elected officers in a certain region, can be sequenced to deliver more information about election results in the region.
Streams for different polls can be overlaid to support synchronized comparison.
\emph{Erie} supports expressing data-driven repetition and concatenation-based composition. 

\subsubsection{Data-driven repetition: Repeat channel}
Data analysts commonly examine a measure conditional on one or more categorical variables.
For instance, the developer may want to extend the walkthrough case by replicating the auditory histogram by the \al{origin} of manufacture (\ie~three histograms for U.S.A., Japan, and Europe). 
To support such cases, a \al{repeat} channel defines how to repeat a \al{stream} design by one or more data fields. 
From the previous example, the developer can repeat the auditory histogram by the \al{origin} and the number of \al{cylinders} (values: 3, 4, 5, 6, and 8):
\begin{align*}
\textit{repeat} = \{ \textit{field} = \eqNValue{[ \textit{origin}, \textit{cylinders} ]} \}
\end{align*}
In this case, the repeat order is nested, such that the histograms for the cylinder values are played for each origin.
A \al{repeat} channel has a \al{speech} property to indicate whether to speak out the value(s) for each repeated stream. 
If \al{speech} is set to \al{true} for this example, the repeated streams are played as shown in \autoref{tab:grammar:repeat1}.

\begin{table}[h]
\caption{The sonification stream order for the auditory histograms repeated by the \al{origin} and \al{cylinders} variables.}\label{tab:grammar:repeat1}
\Description{This is a sonification queue for the auditory histogram repeated by the origin and cylinder variables to demonstrate the case shown in Section 5.6.1. There are 30 queue items while 5 to 10, 13 to 18, and 23 to 28 are omitted and replaced by ellipsis. Queue number 1. Speech. "U.S.A., 3." indicating the origin and cylinder value for the first repetition stream. Queue number 2. Tone. "The histogram for origin U.S.A. and 3 cylinders." Queue number 3. Speech. "U.S.A., 4."  Queue number 4. Tone. "The histogram for origin U.S.A. and 4 cylinders." This structure is repeated for "U.S.A., 8", "Japan, 3", "Japan, 8", "Europe, 3", and "Europe, 8".}
\footnotesize
\def\arraystretch{1.15}%
\arrayrulecolor{gray} 
\rowcolors{2}{white}{black!10!}
\noindent\begin{tabular}{|p{1.25em} p{3em} p{20em}|}
\hline
\textbf{\#} & \textbf{Type} & \textbf{Sound} \\ \hline
1 & Speech & U.S.A., 3 \\ \hline
2 & Tone & [The histogram for origin U.S.A and 3 cylinders] \\ \hline
3 & Speech & U.S.A., 4 \\ \hline
4 & Tone & [The histogram for origin U.S.A and 4 cylinders] \\ \hline
$\ldots$ & $\ldots$ & $\ldots$ \\ \hline
9 & Speech & U.S.A., 8 \\ \hline
10 & Tone & [The histogram for origin U.S.A and 8 cylinders] \\ \hline
11 & Speech & Japan, 3 \\ \hline
12 & Tone & [The histogram for origin Japan and 3 cylinders] \\ \hline
$\ldots$ & $\ldots$ & $\ldots$ \\ \hline
19 & Speech & Japan, 8 \\ \hline
20 & Tone & [The histogram for origin Japan and 8 cylinders] \\ \hline
21 & Speech & Europe, 3 \\ \hline
22 & Tone & [The histogram for origin Europe and 3 cylinders] \\ \hline
$\ldots$ & $\ldots$ & $\ldots$ \\ \hline
29 & Speech & Europe, 8 \\ \hline
30 & Tone & [The histogram for origin Europe and 8 cylinders] \\ \hline
\end{tabular}
\end{table}

Suppose the developer now wants to simultaneously play (\ie~overlay) the auditory histograms for different cylinder values to reduce the playtime.
To do so, the developer can use the \al{by} property in the \al{repeat} channel:
\begin{align*}
\textit{repeat} = \{ & \textit{field} = \eqNValue{[ \textit{origin}, \textit{cylinders} ]}, \textit{by} = \eqNValue{[ \textit{sequence}, \textit{overlay} ]}, \\
& \textit{speech} =  \eqValue{true}\}
\end{align*}
This results in a sonification queue shown in \autoref{tab:grammar:repeat2}.

\begin{table}[h]
\caption{The sonification stream order for the auditory histograms sequenced by the \al{origin} field and overlaid by the \al{cylinders} field.}\label{tab:grammar:repeat2}
\Description{This is a sonification queue for the auditory histogram sequenced by the origin field and overlaid by the cylinder field to demonstrate the case shown in Section 5.6.1. Queue number 1. Speech for "U.S.A." Queue nubmer 2. Tone overlay for "the histograms for U.S.A. and cylinder values of 3, 4, 5, 6, and 8." The same structure repeats for "Japan" and "Europe".}
\footnotesize
\def\arraystretch{1.15}%
\arrayrulecolor{gray} 
\rowcolors{2}{white}{black!10!}
\noindent\begin{tabular}{|p{1.25em} p{6em} p{20em}|}
\hline
\textbf{\#} & \textbf{Type} & \textbf{Sound} \\ \hline
1 & Speech & U.S.A. \\ \hline
2 & Tone-Overlay & 
[The histograms for U.S.A. and cylinder values:]
\begin{tabular}{@{}|l|l|l|l|l|@{}}
\hline
3 & 4 & 5 & 6 & 8 \\
\end{tabular}
\\ \hline
3 & Speech & Japan \\ \hline
4 & Tone-Overlay & 
[The histograms for Japan and cylinder values:]
\begin{tabular}{@{}|l|l|l|l|l|@{}}
\hline
3 & 4 & 5 & 6 & 8 \\
\end{tabular}
\\ \hline
5 & Speech & Europe \\ \hline
6 & Tone-Overlay & 
[The histograms for Europe and cylinder values:]
\begin{tabular}{@{}|l|l|l|l|l|@{}}
\hline
3 & 4 & 5 & 6 & 8 \\
\end{tabular}
\\ \hline
\end{tabular}
\end{table}


\subsubsection{Concatenation: Sequence and overlay}
Two or more separate \al{stream}s can be combined as a \al{sequence} (playing one after another) or an \al{overlay} (playing all together at the same time). 
To enable these multi-stream compositions, we extend the definition of a stream:
\begin{equation*}
\textit{stream} := \{ \textit{data}, \textit{tone}, \textit{encoding}, {\color{blue} \textit{title}, \textit{description}, \textit{config}} \}
\end{equation*}
Consequently, a top-level spec is also redefined as:
\begin{align*}
\textit{spec} := \{ & {\color{blue}(\textit{stream}~|~\textit{overlay}~|~\textit{sequence}), \textit{transform}}, \\ 
& \textit{dataset}, \textit{tick}, \textit{synth}, \textit{wave}, \textit{sampling}, \\
& \textit{title}, \textit{description}, \textit{config} \}
\end{align*}
These extensions allow for specifying the title, description, and configuration of each sub-\al{stream} as well as global data transforms.
The \al{config} object in a sub-\al{stream} overrides the top-level \al{config}.
The \al{transform} object defined in a \al{stream} of a \al{sequence} is applied after the top-level (global) \al{transform} object.

Then, an \al{overlay} is formalized as a list of streams, and a \al{sequence} is defined as an ordered list of streams and overlays:
\begin{align*}
\textit{overlay} & := [ \textit{stream} ] \\
\textit{sequence} & := [\textit{stream} ~|~ \textit{overlay} ]
\end{align*}
Note that a nested sequence, $[\textit{sequence}, \textit{sequence}]$, is also a \al{sequence}.

\subsection{Configuration}\label{sec:grammar:config}
A \al{config} object specifies overall controls for the sonification. 
For example, when a sonification consists of multiple streams that use the same auditory encodings and scales, the developer can skip playing the scale descriptions for the non-first streams.
When a sonification needs more musical representation, a developer can change the \al{time-unit} from seconds (default) to beats.
For background, when BPM is 100, one beat corresponds to 0.6 seconds ($= 60/100$). 
In this case, the developer can specify the tempo (beat per minute, or BPM) and whether to round raw beats to integer beats (\eg~3.224 to 3).
When the time unit of sonification is set to beats, then other time-related units are also accordingly converted. 
For instance, the unit for a \al{tap-speed} channel becomes taps per beat.
\section{\emph{Erie} Compiler and Player for Web}\label{sec:compiler}
A family of compilers and renderers for declarative grammar produces the output as expressed in a design spec. 
For \emph{Erie}, a \textit{queue compiler} compiles a spec to an \textit{audio queue} representing a schedule of sounds to be played in terms of their physical values.
Then, a \textit{player} renders this audio queue into actual sounds. 
We separate the queue compiler from the player to allow listeners to control when to play or pause a sonification and to support developing players for different audio environments, such as CSound~\cite{csound}.
We implemented and open-sourced a spec API, a queue compiler, and a player for a web environment\footnote{\url{https://github.com/see-mike-out/erie-web}} using web standard APIs in JavaScript (\textbf{C5: Compatibility with standards}).

\subsection{Supported Presets}\label{sec:compiler:preset}
Compilers and renderers of declarative grammar often provide default presets.
\emph{Erie} compiler and player offer the following presets.

\bpstart{Data and data transform}
\emph{Erie}'s compiler supports multidimensional data in a relational table form (\eg~CSV, JSON).
Since we assume that a developer has done fundamental data processing and transforms (\eg~fitting a regression model), our compiler supports a minimum set of data transform types that include aggregation, binning, kernel density estimation, folding (columns to rows; \eg~$[\{A: 1, B: 2\}] \rightarrow [\{key: A, value: 1\}, \{key: B, value: e\}]$), filtering, and calculation. 

\bpstart{Instrument types}
Our web player supports musical instruments (classical piano, electronic piano, violin, guitar, metal guitar, clap, hi-hat, high-kick, low-kick), noises (white, pink, and brown), simple oscillators (sine, sawtooth, triangle, and square forms), configurable FM and AM synths, and periodic waves.

\bpstart{Audio filters}
Our web player offers preset filters such as a dynamic compressor, a distortion filter, an envelope node, and various types of biquad filters.
These filters can be chained in the \al{tone} of a \al{stream}.

\bpstart{Encoding channels}
Our queue compiler handles \al{time}, \al{time2}, \al{duration}, \al{tap-speed}, \al{tap-count}, \al{pitch}, \al{detune}, \al{pan}, \al{loudness}, 
\al{timbre}, \al{post-reverb}, \al{modulation} index, 
\al{harmonicity}, \al{speech-before},
\al{speech-after}, and \al{repeat} channels.
Different audio filters can have extra encoding channels. 
For example, a lowpass biquad filter attenuates frequencies above a certain cutoff, and it can have a \al{biquad-frequency} channel to set the cutoff.

\bpstart{Scale descriptions}
\emph{Erie}'s queue compiler generates a description of each scale to give an overview of the sonification.
A scale description functions as an auditory legend in a sonification.
For example, the scales of the \al{time} and \al{pitch} channels from the walkthrough is auditorily described as shown in \autoref{tab:compiler:scale}.

\begin{table}[h]
\caption{The default scale description provided by \emph{Erie} for the walkthrough case. These items are played before the sonification in \autoref{tab:grammar:walkthrough} by default.}\label{tab:compiler:scale}
\Description{The sonification queue for scale description introduced in Section 6.1. The queue numbers start from 4. Queue number 4. Speech for "The miles-per-gallon in mapped to time. The duration of the stream is 4.5 seconds." Queue number 5. Speech for "The count is mapped to pitch. The minimum domain value 0 is mapped to".
Queue number 6. Tone for a sound that starts at time 0, lasts for the duration of 0.3 seconds with a sine oscillator, pitch of 220 Hertz, and loudness of 1. Queue number 7. Speech for "and the maximum domain value 100 is mapped to". Queue number 8. Tone for a sound that starts at time 0, lasts for the duration of 0.3 seconds with a sine oscillator, pitch of 660 Hertz, and loudness of 1.}
\footnotesize
\def\arraystretch{1.15}%
\arrayrulecolor{gray} 
\rowcolors{2}{white}{black!10!}
\noindent\begin{tabular}{|p{1.25em} p{3em} p{24em}|}
\hline
\textbf{\#} & \textbf{Type} & \textbf{Sound} \\ \hline
4 & Speech & The \textit{miles-per-gallon} is mapped to time. The duration of the stream is 4.5 seconds. \\ \hline
5 & Speech & The \textit{count} is mapped to pitch. The minimum domain value 0 is mapped to \\ \hline
6 & Tone & 
\begin{tabular}{@{}p{3.5em}llll@{}}
\qcell{Start}{0} & \qcell{Duration}{0.3} & \qcell{Timbre}{Sine} & \qcell{Pitch}{220} & \qcell{Loudness}{1} \\
\end{tabular}
\\ \hline
7 & Speech & and the maximum domain value 100 is mapped to \\ \hline
8 & Tone & 
\begin{tabular}{@{}p{3.5em}llll@{}}
\qcell{Start}{0} & \qcell{Duration}{0.3} & \qcell{Timbre}{Sine} & \qcell{Pitch}{660} & \qcell{Loudness}{1} \\
\end{tabular}
\\ \hline
\end{tabular}
\end{table}

\subsection{Spec API}\label{sec:compiler:specAPI}
We implemented \emph{Erie} syntax in JavaScript.
For example, the spec of the walkthrough can be written as below.

\begin{lstlisting}[name=apiuse]
// Create a spec object as a single stream.
let spec = new Stream();
// Assign the data URL to the spec.
spec.data("url", "cars.json");
// Add the bin transform
let bin = new Bin("miles-per-gallon"); 
bin
  .as("miles-per-gallon-bin", "miles-per-gallon-bin-end")
  .nice(true); // as/end names -> "auto" binnig
spec.transform.add(bin);
// Add the count aggregation
let aggregate = new Aggregate(); 
// setting operation and the new field name -> setting group-by
aggregate.add("count", "count")
          .groupby(["miles-per-gallon"]);
spec.transform.add(aggregate);
// Set the tone of the stream.
spec.tone.continued(false);
// encodings
// Set the time channel for the "quantitative" field "miles-per-gallon-bin".
// Set the timing to absolute.
spec.encoding.time
  .field("miles-per-gallon-bin", "quantitative")
  .scale("timing", "absolute").scale("length", 4.5);
// Set the time2 channel for the field "miles-per-gallon-bin-end".
spec.encoding.time2.field("miles-per-gallon-bin-end");
// Set the pitch channel for the "quantitative" field "count".
spec.encoding.pitch.field("count", "quantitative")
                    .scale("domain", [0, 100])
                    .scale("range", [220, 660])
                    .scale("polarity", "positive");
\end{lstlisting}

\noindent This spec is equivalent to the following JSON object, which can be obtained via the \code{get} method of the spec API.
This JSON syntax reuses some Vega-Lite~\cite{satyanarayan:vega-lite2017} expressions, supporting cases where visualization and sonification need to be provided concurrently.

\begin{lstlisting}[name=apiuse]
// results of spec.get()
{ "data": { "url": "cars.json" },
  "transform": [{
    "bin": "miles-per-gallon",
    "as": "miles-per-gallon-bin",
    "end": "miles-per-gallon-bin-end",
    "nice": true,
  }, {
    "aggregate": [{ "op": "count", "as": "count" }],
    "groupby": ["miles-per-gallon-bin"] } ],
  "tone": { "continued": false },
  "encoding": {
    "time": {
      "field": "miles-per-gallon-bin",
      "type": "quantitative",
      "scale": { "timing": "absolute", "length": 4.5 } }, 
    "time2": { "field": "miles-per-gallon-bin-end" }
    "pitch": {
      "field": "count",
      "type": "quantitative",
      "scale": { "domain": [0, 100], "range": [220, 660] } } } }
\end{lstlisting}

\subsection{Queue Compiler}\label{sec:compiler:queue}
Given a spec, our queue compiler converts data values to auditory values.
The outcome audio queue is an ordered list of sub-queues, and each sub-queue item can have one of these four types: \textit{speech}, \textit{tone-series}, \textit{tone-speech-series}, and \textit{tone-overlay}. 
A \textit{speech} queue consists of natural language sentences that are played one after another.
A \textit{tone-series} queue is a timed list of non-speech sounds, and a \textit{tone-speech-series} queue is a timed list of sounds and speeches. 
Each sound in a sub-queue of these two types is expressed in terms of their actual auditory values (\eg~Hz for pitch).
Lastly, a \textit{tone-overlay} queue consists of multiple \textit{tone-series} queues that are played simultaneously. 
An audio queue is not nested except \textit{tone-overlay} queues, and a \al{sequence} spec is compiled to multiple flattened sub-queues.

To compile a spec into an audio queue, a developer can run \code{compileAudioGraph} function, which asynchronously compiles the spec to an audio queue:
\begin{lstlisting}[name=apiuse]
let audioQueue = await compileAuidoGraph(spec.get());
\end{lstlisting}

\subsection{Player for Web}\label{sec:compiler:player}
We developed an \emph{Erie} player for web environments using the standard Web Audio API~\cite{webAudio} and Web Speech API~\cite{webSpeech}.
The player offers several playing options: play from the beginning, pause, resume, stop, play from a sub-queue, and play from one sub-queue to another.
\begin{lstlisting}[name=apiuse]
audioQueue.queue.play(); // Play from the beginning
audioQueue.queue.pause(); // Pause
audioQueue.queue.resume(); // Resume from where it was paused
audioQueue.queue.stop(); // Stop playing
audioQueue.queue.play(i); // Play from the i-th sub-queue
audioQueue.queue.play(i, j); // Play the i-th to (j-1)-th sub-queues.
\end{lstlisting}

\subsection{Filter and Channel Extension}\label{sec:compiler:extension}
To achieve certain sound effects, a developer could use audio filters in addition to custom instruments (\eg~configured synth, sampling).
Furthermore, those audio filters can encode data variables (\eg~the amount of distortion to express air quality).
To widen such design possibilities, \emph{Erie} offers APIs for defining custom audio filters that can have additional encoding channels (\textbf{C4: Extensibility}).

To describe the process of defining a custom filter, imagine that a developer wants to add an envelope filter with encodable \code{attack} and \code{release} times. 
\code{Attack} means the time duration from the zero volume at the beginning of a sound to the highest volume, and \code{release} refers to the time taken from the highest volume to the zero volume at the end of the sound~\cite{appleEnvelope}.
The developer first needs to define the filter as a JavaScript \code{class} that can be chained from a sonification sound to an output audio device.
This class should have \code{connect} and \code{disconnect} methods to enable the chaining, following the Web Audio API syntax~\cite{MDN:audionode}.
Then, the developer needs to define an \code{encoder} function that assigns the \code{attack} value for each data value to the filter and a \code{finisher} function that assigns the \code{release} values to the filter. 
Refer to the documentation in our Supplementary Material for technical details.

\section{Demonstration}\label{sec:demo}
To demonstrate \emph{Erie} grammar's \textbf{independence from visualization (C1)} and \textbf{expressiveness (C2)}, we walk through novel examples.
We also replicated and extended prior sonifications to show the feasibility of our compiler and player for sonification development.
In addition to the below examples, more use cases, such as a confidence interval, histogram, and sonification of COVID-19 death tolls, are available in our example gallery\footnote{\url{https://see-mike-out.github.io/erie-editor/}}.

\subsection{Example Sonification Designs}\label{sec:demo:example}
We show three representative example cases to show how \emph{Erie} can be used.

\subsubsection{Data sparsity}\label{sec:demo:sparsity}
Given five data tables named A to E, suppose we want to identify and compare their sparsity (the portion of cells that are empty) using a tap-speed channel. 
We have a nominal variable, dataset \code{name}, and a quantitative variable, \code{sparsity}, and the data looks like:
\begin{lstlisting}[name=sparsity]
let data = [
 { "name": "A", "sparsity": 0.4 },
 { "name": "B", "sparsity": 0.6 },
 { "name": "C", "sparsity": 0.2 },
 { "name": "D", "sparsity": 0 },
 { "name": "E", "sparsity": 0.9 }];
\end{lstlisting}

Now, we want to map the \code{name} field to the \code{time} channel of a sonification and the \code{sparsity} to the \code{tapSpeed} channel, so that a sparse dataset with a higher sparsity value has slower tapping.
First, we create a single-stream sonification spec object and set a description text.
\begin{lstlisting}[name=sparsity]
let spec = new Stream();
spec.description("The sparsity of different datasets.");
\end{lstlisting}
Then, we assign the \code{data} to this spec.
\begin{lstlisting}[name=sparsity]
spec.data("values", data);
\end{lstlisting}
With a default sine-wave oscillator, we need a discrete tone to represent separate data tables, which can be specified as: 
\begin{lstlisting}[name=sparsity]
spec.tone.type("default").continued(false);
\end{lstlisting}

\begin{table}[t]
\caption{The audio queue resulting from a sparsity sonification spec in \autoref{sec:demo:sparsity}. 
``Q'' indicates the index of each sub-queue.
``After prev.'' means ``play after the previous sound'' within the same sub-queue.
A tapping pattern, $[a, b] \times c$, means a tap sound for $a$ seconds and a pause for $b$ seconds are repeated $c$ times (the last pause is omitted).
A tapping pattern, $[a, b, c]$, means a pause for $a$ seconds, a tap sound for $b$ seconds, and a pause for $c$ seconds.}\label{tab:demo:sparsity}
\Description{The sonification queue for a sparsity sonification in Section 7.1.1. It has eleven subqueue items.}
\footnotesize
\def\arraystretch{1.15}%
\arrayrulecolor{gray} 
\rowcolors{2}{white}{black!10!}
\noindent\begin{tabular}{|p{1.25em} p{3em} p{25.5em}|}
\hline
\textbf{Q.} & \textbf{Type} & \textbf{Sound} \\ \hline
1 & Speech & To stop playing the sonification, press the X key. \\ \hline
2 & Speech & The sparsity of different datasets. \\ \hline
3 & Speech & This stream has the following sound mappings. \\ \hline
4 & Speech & The category is mapped to time. \\ \hline
5 & Speech & The sparsity is mapped to tap speed. The minimum value 0 is mapped to \\ \hline
6 & Tone & 
\begin{tabular}{@{}p{3.4em}lllll@{}}
\tcell{Start}{0}{} & \tcell{Dur.}{2}{} & \tcell{Timbre}{Sine}{} & \tcell{Pitch}{523.25 (C5)}{} & \tcell{Loud.}{1}{} & \tcell{Tapping}{$[.19, .01] \times 10$}{\includegraphics[height=0.475em]{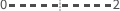}}  \\
\end{tabular}
\\ \hline
7 & Speech & and the maximum value 1 is mapped to. \\ \hline
8 & Tone & 
\begin{tabular}{@{}p{3.4em}lllll@{}}
\qcell{Start}{0} & \qcell{Dur.}{2} & \qcell{Timbre}{Sine} & \qcell{Pitch}{523.25 (C5)} & \qcell{Loud.}{1} & \qcell{Tapping}{No tapping}  \\
\end{tabular}
\\ \hline
9 & Speech & Start playing. \\ \hline
10 & {Tone-Speech} & 
\makecell[l]{
\begin{tabular}{@{}p{3.4em}ll@{}}
\qcell{Start}{0} & \qcell{Dur.}{-} & \qcell{Speech}{``A''} \\ 
\end{tabular} \\ 
\begin{tabular}{@{}p{3.4em}lllll@{}}
\hline
\tcell{Start}{After prev.}{} & \tcell{Dur.}{2}{} & \tcell{Timbre}{Sine}{} & \tcell{Pitch}{523.25 (C5)}{} & \tcell{Loud.}{1}{} & \tcell{Tapping}{$[.19,.17] \times 6$}{\includegraphics[height=0.475em]{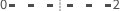}} \\
\hline
\end{tabular} \\
\begin{tabular}{@{}p{3.4em}ll@{}}
\qcell{Start}{After prev.} & \qcell{Dur.}{-} & \qcell{Speech}{``B''} \\ 
\end{tabular} \\ 
\begin{tabular}{@{}p{3.4em}lllll@{}}
\hline
\tcell{Start}{After prev.}{} & \tcell{Dur.}{2}{} & \tcell{Timbre}{Sine}{} & \tcell{Pitch}{523.25 (C5)}{} & \tcell{Loud.}{1}{} & \tcell{Tapping}{$[.19, .41] \times 4$}{\includegraphics[height=0.475em]{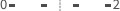}} \\
\hline
\end{tabular} \\
\begin{tabular}{@{}p{3.4em}ll@{}}
\qcell{Start}{After prev.} & \qcell{Dur.}{-} & \qcell{Speech}{``C''} \\ 
\end{tabular} \\ 
\begin{tabular}{@{}p{3.4em}lllll@{}}
\hline
\tcell{Start}{After prev.}{} & \tcell{Dur.}{2}{} & \tcell{Timbre}{Sine}{} & \tcell{Pitch}{523.25 (C5)}{} & \tcell{Loud.}{1}{} & \tcell{Tapping}{$[.19, .07] \times 8$}{\includegraphics[height=0.475em]{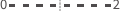}} \\
\hline
\end{tabular} \\
\begin{tabular}{@{}p{3.4em}ll@{}}
\qcell{Start}{After prev.} & \qcell{Dur.}{-} & \qcell{Speech}{``D''} \\ 
\end{tabular} \\ 
\begin{tabular}{@{}p{3.4em}lllll@{}}
\hline
\tcell{Start}{After prev.}{} & \tcell{Dur.}{2}{} & \tcell{Timbre}{Sine}{} & \tcell{Pitch}{523.25 (C5)}{} & \tcell{Loud.}{1}{} & \tcell{Tapping}{$[.19, .01] \times 10$}{\includegraphics[height=0.475em]{figures/sp_tap_4.pdf}} \\
\hline
\end{tabular} \\
\begin{tabular}{@{}p{3.4em}ll@{}}
\qcell{Start}{After prev.} & \qcell{Dur.}{-} & \qcell{Speech}{``E''} \\ 
\end{tabular} \\
\begin{tabular}{@{}p{3.4em}lllll@{}}
\hline
\tcell{Start}{After prev.}{} & \tcell{Dur.}{2}{} & \tcell{Timbre}{Sine}{} & \tcell{Pitch}{523.25 (C5)}{} & \tcell{Loud.}{1}{} & \tcell{Tapping}{$[.91, .19, .91]$}{\includegraphics[height=0.475em]{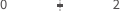}} \\
\end{tabular}
}
\\ \hline
11 & Speech & Finished. \\ \hline
\end{tabular}
\end{table}

Next, we set the \code{time} encoding channel as described earlier.
\begin{lstlisting}[name=sparsity]
spec.encoding.time.field("name", "nominal");
\end{lstlisting}
This \code{time} channel should use relative timing to allow for playing each data table name before the sound for the corresponding sparsity value.
\begin{lstlisting}[name=sparsity]
spec.encoding.time.scale("timing", "relative");
\end{lstlisting}
We then specify the \code{tapSpeed} channel for the quantitative sparsity channel.
\begin{lstlisting}[name=sparsity]
spec.encoding.tapSpeed.field("sparsity", "quantitative");
\end{lstlisting}
This \code{tapSpeed} channel has the domain of $[0, 1]$. We want to map this domain to the range of $[0, 5]$ (\ie~zero to five taps per second) for 2 seconds:
\begin{lstlisting}[name=sparsity]
spec.encoding.tapSpeed.scale("domain", [0, 1])
  .scale("range", [0, 5]).scale("band", 2);
\end{lstlisting}
Since a higher sparsity value should have a lower speed, we need negative \code{polarity}:
\begin{lstlisting}[name=sparsity]
spec.encoding.tapSpeed.scale("polarity", "negative");
\end{lstlisting}
This results in a single tap sound for the sparsity value of $0.1$.
To play this sound in the middle of the time \code{band} (two seconds), we set the \code{singleTappingPosition} property as \code{middle}:
\begin{lstlisting}[name=sparsity]
spec.encoding.tapSpeed
  .scale("singleTappingPosition", "middle");
\end{lstlisting}
To support identifying these tapping sounds at different speeds, we need a \code{speechBefore} channel for the \code{name} channel. 
\begin{lstlisting}[name=sparsity]
spec.encoding.speechBefore.field("name", "nominal");
\end{lstlisting}
We do not need a scale description for this \code{speechBefore} channel in this case.
\begin{lstlisting}[name=sparsity]
spec.encoding.speechBefore.scale("description", "skip");
\end{lstlisting}
\autoref{tab:demo:sparsity} shows the audio queue compiled from this spec.

\subsubsection{Kernel density estimation}\label{sec:demo:kde}

In exploratory data analysis pipe\-lines, examining the distributions of variables of interest is a common first step.
It is important to observe the entirety of a distribution because some distributional information, such as multi-modality, are not captured by summary statistics like mean and variance.
In addition to histograms, data analysts often estimate the probability density of a quantitative variable using a kernel smoothing function (\ie~kernel density estimation or KDE). 
In this example, we want to sonify a KDE of the \code{bodyMass} variable of the \code{penguins.json} data~\cite{vegadataset}.
This sonification will encode the density by pitch and the variable's value by time and panning.
Then, we repeat this sonification design for different \code{species} and \code{island}s of penguins. 

The \code{penguins.json} dataset consists of \code{species}, \code{island}, and \code{bodyMass} fields.
The nominal \code{species} and \code{island} fields form five combinations as shown in the first two columns of \autoref{tab:demo:kde:data}.
The \code{bodyMass} field roughly ranges from 2,500 to 6,500.

\begin{table}[b]
\caption{A preview of the \code{penguins.json} dataset.}\label{tab:demo:kde:data}
\Description{This table shows a few data points in the penguins.json dataset.}
\def\arraystretch{1.15}%
\arrayrulecolor{lightgray}
    \centering
    \footnotesize
    \begin{tabular}{lll}
    \hline
        \textbf{\code{species}} & \textbf{\code{island}} & \textbf{\code{bodyMass}} \\ \hline
        Adelie & Torgersen & 3,750 \\ \hline
        Adelie & Biscoe & 4,300 \\ \hline
        Adelie & Dream & 2,900 \\ \hline
        Chinstrap & Dream & 3,450 \\ \hline
        Gentoo & Biscoe & 6,300 \\ \hline
    \end{tabular}
\end{table}

First, we create a single-stream spec object, set the description, and assign the data.
\begin{lstlisting}[name=kde]
let spec = new Stream();
spec.description("The kernel density estimation of body mass by species and island.");
spec.data("url", "penguins.json");
\end{lstlisting}

Next, we need to add a \code{density} transform for the KDE of the \code{bodyMass} field grouped by \code{species} and \code{island}.
\begin{lstlisting}[name=kde]
let density = new Density();
density.field("bodyMass").extent([2500, 6500])
  .groupby(["species", "island"]);
spec.transform.add(density);
\end{lstlisting}
This transform results in a new data table that has four columns: \code{value} (evenly distributed \code{bodyMass} values, \eg~2500, 2550, \ldots, 6450, 6500), \code{density} (the density estimate of each \code{value} element), \code{species}, and \code{island}.

Third, we use a \code{continued} tone because we want to sonify continuous KDEs.
\begin{lstlisting}[name=kde]
spec.tone.type("default").continued(true);
\end{lstlisting}

Given this \code{tone} design, we set the \code{time}, \code{pan}, and \code{pitch} channels.
We map the \code{value} field to \code{time} and \code{pan} to give both temporal and spatial senses of sound progression. 
\begin{lstlisting}[name=kde]
spec.encoding.time.field("value", "quantitative");
spec.encoding.pan.field("value", "quantitative");
\end{lstlisting}
Then, we detail the \code{scale} of the \code{time} channel by setting the \code{length} of each repeated sound to three seconds and indicating the \code{title} of this \code{scale} in the scale description. 
\begin{lstlisting}[name=kde]
spec.encoding.time.scale("length", 3)
  .scale("title", "Body Mass values");
\end{lstlisting}
Similarly, we set the \code{polarity} of the \code{pan} channel to \code{positive} and note the same scale \code{title}.
\begin{lstlisting}[name=kde]
spec.encoding.pan.scale("polarity", "positive")
  .scale("title", "Body Mass values");
\end{lstlisting}
We encode the \code{density} field to the \code{pitch} channel with \code{positive} \code{polarity} and a pitch range of \code{[0, 700]} (Hz).
\begin{lstlisting}[name=kde]
spec.encoding.pitch.field("density", "quantitative")
                   .scale("polarity", "positive")
                   .scale("range", [0, 700])
                   .scale("title", "kernel density");
\end{lstlisting}
KD estimates usually have infinite decimals (\eg~$0.0124 \ldots$), which makes it hard to understand when read out (\eg~in the scale description).
To prevent reading all the decimals, we specify the read \code{format} of the density estimates so that they are only read up to the fourth decimal.
\begin{lstlisting}[name=kde]
spec.encoding.pitch.format(".4");
\end{lstlisting}
\emph{Erie} uses format expressions supported by D3.js~\cite{bostock:d32011}.

Now, we repeat this spec design by the \code{species} and \code{island} fields using a \code{repeat} channel.
\begin{lstlisting}[name=kde]
spec.encoding.repeat
  .field(["species", "island"], "nominal")
  .speech(true).scale("description", "skip");
\end{lstlisting}

\autoref{tab:demo:kde} illustrates the audio queue compiled from this spec. Sub-queue 4 to Sub-queue 8 are the scale descriptions for the \code{time}, \code{pan}, and \code{pitch} channels with audio legends. 
Sub-queues 10 to 24 represent the specified KDE sonification for each combination of the \code{species} and \code{island} values.

\begin{table}[t]
\caption{The audio queue resulting from a kernel density estimate sonification spec in \autoref{sec:demo:kde}. 
``Q'' indicates the index of each sub-queue. 
The pitch values (range from 0 to 700) are low because they are representing the both-side tails of each estimated density distribution.}\label{tab:demo:kde}
\Description{This is the sonification queue for a kernel density estimate design in Section 7.1.2. There are 25 subqueue items with 13 to 21 are omitted with an ellipsis.}
\footnotesize
\def\arraystretch{1.15}%
\arrayrulecolor{gray} 
\rowcolors{2}{white}{black!10!}
\noindent\begin{tabular}{|p{1.25em} p{3em} p{25.5em}|}
\hline
\textbf{Q.} & \textbf{Type} & \textbf{Sound} \\ \hline
1 & Speech & To stop playing the sonification, press the X key. \\ \hline
2 & Speech & Kernel density of Body Mass by Species and Island. \\ \hline
3 & Speech & This stream has the following sound mappings. \\ \hline
4 & Speech & The Body Mass value is mapped to time. The duration of the stream is 3 seconds. \\ \hline
5 & Speech & The Body Mass value is mapped to pan. The domains values from 2500 to 6500 are mapped to \\ \hline
6 & Tone & 
\begin{tabular}{@{}p{3.5em}llll@{}}
\qcell{Start}{0} & \qcell{Timbre}{Sine} & \qcell{Pitch}{523.25} & \qcell{Pan}{-1 (leftmost)} & \qcell{Loudness}{1}  \\ \hline
\qcell{Start}{0.6} & \qcell{Timbre}{Sine} & \qcell{Pitch}{523.25} & \qcell{Pan}{1 (rightmost)} & \qcell{Loudness}{1}  \\
\end{tabular}
\\ \hline
7 & Speech & The Kernel density is mapped to pitch. The domains values from 1.654e-30 to 0.0011 are mapped to \\ \hline
8 & Tone & 
\begin{tabular}{@{}p{3.5em}llll@{}}
\qcell{Start}{0} & \qcell{Timbre}{Sine} & \qcell{Pitch}{0} & \qcell{Pan}{0 (center)} & \qcell{Loudness}{1}  \\ \hline
\qcell{Start}{0.6} & \qcell{Timbre}{Sine} & \qcell{Pitch}{700} & \qcell{Pan}{0} & \qcell{Loudness}{1}  \\
\end{tabular}
\\ \hline
9 & Speech & This sonification sequence consists of 5 parts. \\ \hline
10 & Speech & Stream 1. Adelie and Torgersen. \\ \hline
11 & Speech & Start playing. \\ \hline
12 & Tone & 
\begin{tabular}{@{}p{3.5em}llll@{}}
\qcell{Start}{0} & \qcell{Timbre}{Sine} & \qcell{Pitch}{7.3928} & \qcell{Pan}{-1} & \qcell{Loudness}{1}  \\ \hline
\qcell{Start}{0.015} & \qcell{Timbre}{Sine} & \qcell{Pitch}{9.0813} & \qcell{Pan}{-0.99} & \qcell{Loudness}{1}  \\ \hline
\multicolumn{5}{c}{$\cdots$} \\ \hline
\qcell{Start}{2.88} & \qcell{Timbre}{Sine} & \qcell{Pitch}{6.0194} & \qcell{Pan}{0.92} & \qcell{Loudness}{1}  \\ \hline
\qcell{Start}{3} & \qcell{Timbre}{Sine} & \qcell{Pitch}{1.4486} & \qcell{Pan}{1} & \qcell{Loudness}{1}  \\
\end{tabular}
\\ \hline
\multicolumn{3}{c}{$\cdots$} \\ \hline
22 & Speech & Stream 5. Gentoo and Biscoe. \\ \hline
23 & Speech & Start playing. \\ \hline
24 & Tone & 
\begin{tabular}{@{}p{3.5em}llll@{}}
\qcell{Start}{0} & \qcell{Timbre}{Sine} & \qcell{Pitch}{0.0000} & \qcell{Pan}{-1} & \qcell{Loudness}{1}  \\ \hline
\multicolumn{5}{c}{$\cdots$} \\ \hline
\qcell{Start}{3} & \qcell{Timbre}{Sine} & \qcell{Pitch}{9.2425} & \qcell{Pan}{1} & \qcell{Loudness}{1}  \\ 
\end{tabular}
\\ \hline 
25 & Speech & Finished. \\ \hline
\end{tabular}
\end{table}

\subsubsection{Model fit sequence}\label{sec:demo:fit}

After fitting a linear regression model, a necessary task is to check the model fit by examining the residuals.
Common methods for residual analysis include a residual plot (residual vs. independent variable) and a QQ plot (residual vs. normal quantile). 
For this task, we assume that we have already fitted a linear regression model of Sepal Length on Petal Length ($\textit{Petal Length} \sim \textit{Sepal Length}$), and computed the residuals.
For the residual plot, we use a \code{residuals} dataset with two fields: \code{sepalLength} (independent variable) and \code{residuals}. 
For the QQ plot, we use a \code{qq} dataset with two fields: \code{normalQuantile} and \code{residuals}\footnote{Alternatively, these two datasets can be a single dataset. Here, we use two datasets for demonstration purposes.}.
With these datasets, we want to create two sequenced sonifications for residuals and comparison to normal quantiles (\ie~recognizing their trends). 

We first register the datasets.
\begin{lstlisting}[name=fit]
let qqData = [...];
let qqDataset = new Dataset("qq");
qqDataset.set("values", qqData);
let residualData = [...];
let residualDataset = new Dataset("residuals");
residualDataset.set("values", residualData);
\end{lstlisting}

Second, we define a sonification for a residual plot.
When errors of a model fit are unbiased, the residuals are evenly distributed along values of the independent variable and on both sides of the central line indicating 0 error. 
With this residual plot sonification, we want to capture the evenness of residual distribution by mapping the residuals to \code{modulation} index and \code{pan} channel. 
In this way, a larger residual is mapped to a more warped sound, and a negative residual is played on the left side and a positive residual is played on the right side.
A good model fit will generate a sonification where the sound quickly (\eg~150 sound points within 5 seconds) moves between different modulation index and pan values, making it harder to differentiate their degrees of warping and spatial positions. 
In contrast, a bad model fit will generate an audio graph where listeners can easily sense some groups of sounds sharing the same degree of warping on a certain spatial position. 
We use a \code{time} channel for the \code{sepalLength} field.

To do so, we create a single stream with the \code{residuals} dataset.
\begin{lstlisting}[name=fit]
let residualSpec = new Stream();
residualSpec.name("Residuals");
residualSpec.data.set(residualData);
\end{lstlisting}
For the tone, we use an FM synth, named \code{fm1}.
\begin{lstlisting}[name=fit]
let synth = new SynthTone("fm1");
synth.type("FM");
residualSpec.tone.set(synth);
\end{lstlisting}
The residual sonification uses a \code{time} channel for the \code{sepalLength} values and \code{modulation} index and \code{pan} channels for the residuals that roughly range from $-2.5$ to $2.5$.
This design is specified as below:
\begin{lstlisting}[name=fit]
residualSpec.encoding.time
  .field("sepalLength", "quantitative")
  .scale("timing", "absolute").scale("legnth", 5)
  .scale("band", 0.15).format(".4");
residualSpec.encoding.modulation
  .field("residual", "quantitative")
  .scale("domain", [-2.5, 0, 2.5])
  .scale("range", [4, 0.001, 4]).format(".4");
residualSpec.encoding.pan
  .field("residual", "quantitative")
  .scale("domain", [-2.5, 0, 2.5])
  .scale("range", [-1, 0, 1]).format(".4");
\end{lstlisting}

Next, we specify a QQ plot sonification.
A good model fit should also exhibit normally distributed residuals.
By plotting the quantiles of the residuals against the expected quantiles of a normal distribution (range from 0 to 1), we want to observe how much the residuals deviate from the expectation that they are normally distributed.
A visual QQ plot shows the gap between the theoretical and observed distribution by plotting them in a Cartesian space, which is the same format used for a residual plot at high level. 
However, a sonification author may need to directly encode the gap because overlaying the normal and residual distributions with different pitches or volumes may make it harder to decode the gap, indicating the need for specifying a sonification design \textbf{independently from visualization (C1)}. 
Thus, we compute the normalized residuals' \code{deviation} (within 0 to 1) from normal quantiles to directly encode the gap. 
This transform is done using the below \code{calculate} transforms, resulting in two additional fields: \code{normalized} and \code{deviation}.
\begin{lstlisting}[name=fit]
let qqSpec = new Stream();
qqSpec.name("QQ plot");
qqSpec.data.set(residualData);
// normalize residuals using its minimum (-2.477468) and maximum (2.495122).
let noramlized = new Calculate("(datum.residual + 2.477468)/(2.495122 + 2.477468)", "normalized");
let deviation = new Calculate("datum.normalized - datum.normalQuantile", "deviation");
qqSpec.transform.add(normalized).add(deviation);
\end{lstlisting}

Then, we map the \code{normalQuantile} to \code{time}, the magnitude of the \code{residual} to \code{pitch}, and the \code{deviation} to \code{pan}.
These mappings will produce sounds that are spatially centered when the deviation is smaller but are played from left or right when the signed deviation is bigger.
\begin{lstlisting}[name=fit]
qqSpec.tone.continued(false);
qqSpec.encoding.time
  .field("normalQuantile", "quantitative")
  .scale("length", 4).scale("band", 0.2)
  .scale("title", "Normal Quantile").format(".4");
qqSpec.encoding.pitch
  .field("residual", "quantitative")
  .scale("polarity", "positive")
  .scale("title", "Residual").format(".4");
qqSpec.encoding.pan
  .field("deviation", "quantitative")
  .scale("domain", [-0.2, 0, 0.2])
  .scale("range", [-1, 0, 1])
  .scale("title", "Deviation from normal distribution")
  .format(".4");
\end{lstlisting}

Lastly, we merge the residual and QQ streams (\code{residualSpec}, \code{qqSpec}) into a sequenced stream.
\begin{lstlisting}[name=fit]
let modelFit = new Sequence(residualSpec, qqSpec);
modelFit.description("The residuals of a linear regression model of Sepal Length on Petal Length.");
\end{lstlisting}
This spec results in the sonification described in \autoref{tab:demo:fit}.

\begin{table}[!h]
\caption{The audio queue resulting from a model fit sonification spec in \autoref{sec:demo:fit}. 
``Q'' indicates the index of each sub-queue.}\label{tab:demo:fit}
\Description{This is the sonification queue for the model fit in Section 7.1.3. There are 36 subqueue items.}
\footnotesize
\def\arraystretch{1.15}%
\arrayrulecolor{gray} 
\rowcolors{2}{white}{black!10!}
\noindent\begin{tabular}{|p{1.15em} p{3em} p{26em}|}
\hline
\textbf{Q.} & \textbf{Type} & \textbf{Sound} \\ \hline
1 & Speech & To stop playing the sonification, press the X key. \\ \hline
2 & Speech & The residuals of a linear regression model of Sepal Length on Petal Length. \\ \hline
3 & Speech & This sonification sequence consists of 2 parts. \\ \hline
4 & Speech & Stream 1. Residual plot. \\ \hline
5 & Speech & The first stream has the following sound mappings. \\ \hline
6 & Speech & The Sepal Length is mapped to time. The duration of the stream is 5 seconds. \\ \hline
7 & Speech & The residual is mapped to pan. Residual values are mapped as -2.5  \\ \hline
8 & Tone & 
\begin{tabular}{@{}llllll@{}}
\qcell{Start}{0} & \qcell{Duration}{0.3} & \qcell{Timbre}{fm1} & \qcell{Pitch}{523.25} & \qcell{Loudness}{1}  & \qcell{pan}{-1} \\
\end{tabular}
\\ \hline
9 & Speech & 0 (zero) \\ \hline
10 & Tone & 
\begin{tabular}{@{}llllll@{}}
\qcell{Start}{0} & \qcell{Duration}{0.3} & \qcell{Timbre}{fm1} & \qcell{Pitch}{523.25} & \qcell{Loudness}{1}  & \qcell{pan}{0} \\
\end{tabular} \\ \hline
11 & Speech & 2.5  \\ \hline
12 & Tone & 
\begin{tabular}{@{}llllll@{}}
\qcell{Start}{0} & \qcell{Duration}{0.3} & \qcell{Timbre}{fm1} & \qcell{Pitch}{523.25} & \qcell{Loudness}{1}  & \qcell{pan}{1} \\
\end{tabular} \\  \hline
13 & Speech & The residual is mapped to modulation. Residual values are mapped as -2.5  \\ \hline
14 & Tone & 
\begin{tabular}{@{}llllll@{}}
\qcell{Start}{0} & \qcell{Duration}{0.3} & \qcell{Timbre}{fm1} & \qcell{Pitch}{523.25} & \qcell{Loudness}{1}  & \qcell{Modulation}{4} \\
\end{tabular}
\\ \hline
15 & Speech & 0 (zero) \\ \hline
16 & Tone & 
\begin{tabular}{@{}llllll@{}}
\qcell{Start}{0} & \qcell{Duration}{0.3} & \qcell{Timbre}{fm1} & \qcell{Pitch}{523.25} & \qcell{Loudness}{1}  & \qcell{Modulation}{0.001} \\
\end{tabular} \\ \hline
17 & Speech & 2.5  \\ \hline
18 & Tone & 
\begin{tabular}{@{}llllll@{}}
\qcell{Start}{0} & \qcell{Duration}{0.3} & \qcell{Timbre}{fm1} & \qcell{Pitch}{523.25} & \qcell{Loudness}{1}  & \qcell{Modulation}{4} \\
\end{tabular} \\  \hline
19 & Speech & Start playing. \\ \hline
20 & Tone & 
\begin{tabular}{@{}lllllll@{}}
\qcell{Start}{0} & \qcell{Duration}{0.15} & \qcell{Timbre}{fm1} & \qcell{Pitch}{523.25} & \qcell{Loud.}{1} & \qcell{Pan}{0.0841} & \qcell{Modulation}{0.3372} \\  \hline
\multicolumn{6}{c}{$\cdots$} \\ \hline
\qcell{Start}{4.85} & \qcell{Duration}{0.15} & \qcell{Timbre}{fm1} & \qcell{Pitch}{523.25} & \qcell{Loud.}{1} & \qcell{Pan}{-0.4721}  & \qcell{Modulation}{1.8888} \\
\end{tabular} \\  \hline
21 & Speech & Stream 2. QQ plot. \\ \hline
22 & Speech & The second stream has the following sound mappings. \\ \hline
23 & Speech & The Normal Quantile is mapped to time. The duration of the stream is 4 seconds. \\ \hline
24 & Speech & The Deviation from normal distribution is mapped to pan. Deviation from normal distribution values are mapped as -0.2 \\ \hline
25 & Tone & 
\begin{tabular}{@{}llllll@{}}
\qcell{Start}{0} & \qcell{Duration}{0.3} &  \qcell{Timbre}{Sine} & \qcell{Pitch}{523.25} & \qcell{Loudness}{1}  & \qcell{Pan}{-1} \\
\end{tabular}
\\ \hline
26 & Speech & 0 (zero) \\ \hline
27 & Tone & 
\begin{tabular}{@{}llllll@{}}
\qcell{Start}{0} & \qcell{Duration}{0.3} &  \qcell{Timbre}{Sine} & \qcell{Pitch}{523.25} & \qcell{Loudness}{1}  & \qcell{Pan}{0} \\
\end{tabular}
\\ \hline
28 & Speech & 0.2  \\ \hline
29 & Tone & 
\begin{tabular}{@{}llllll@{}}
\qcell{Start}{0} & \qcell{Duration}{0.3} &  \qcell{Timbre}{Sine} & \qcell{Pitch}{523.25} & \qcell{Loudness}{1}  & \qcell{Pan}{1} \\
\end{tabular}
\\ \hline
30 & Speech & The Residual is mapped to pitch. The minimum value -2.477 is mapped to \\ \hline
31 & Tone & 
\begin{tabular}{@{}lllll@{}}
\qcell{Start}{0} & \qcell{Duration}{0.3} &  \qcell{Timbre}{Sine} & \qcell{Pitch}{207.65} & \qcell{Loudness}{1} \\
\end{tabular}
\\ \hline
32 & Speech & and the maximum value 2.495 is mapped to \\ \hline
33 & Tone & 
\begin{tabular}{@{}lllll@{}}
\qcell{Start}{0} & \qcell{Duration}{0.3} &  \qcell{Timbre}{Sine} & \qcell{Pitch}{1600} & \qcell{Loudness}{1}  \\
\end{tabular}
\\ \hline
34 & Speech & Start playing. \\ \hline
35 & Tone & 
\begin{tabular}{@{}llllll@{}}
\qcell{Start}{0} & \qcell{Duration}{0.2} &  \qcell{Timbre}{Sine} & \qcell{Pitch}{207.65} & \qcell{Loudness}{1}  & \qcell{Pan}{-0.0167} \\  \hline
\multicolumn{6}{c}{$\cdots$} \\ \hline
\qcell{Start}{3.8} & \qcell{Duration}{0.2} &  \qcell{Timbre}{Sine} & \qcell{Pitch}{1600} & \qcell{Loudness}{1}  & \qcell{Pan}{0.0167} \\
\end{tabular} \\ \hline
36 & Speech & Finished. \\ \hline
\end{tabular}
\end{table}

\subsection{Replication of Prior Use Cases}\label{sec:replication}
We replicate several sonification use cases (\eg~applications and data stories) and extend their features to demonstrate how feasibly creators can use \emph{Erie} in development settings.
We include the \emph{Erie} specs used for the below replications in our example gallery\footnote{\url{https://see-mike-out.github.io/erie-editor/}}.

\subsubsection{Audio Narrative}\label{sec:demo:narrative}

\begin{figure}
    \centering
    \includegraphics[width=\columnwidth]{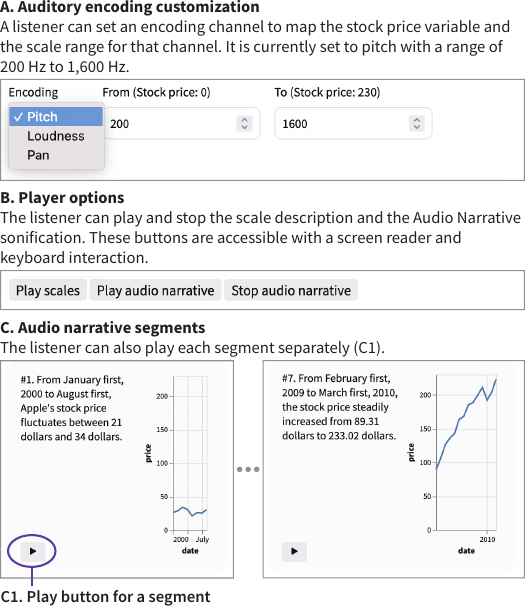}
    \caption{Our replication and extension of Audio Narrative~\cite{audioNarrative:siu2022} using \emph{Erie}. In addition to the originally offered sequencing and speech description, we included options for using different encoding channels (A) and playing the scale description (B).}
    \Description{This figure has three parts from A to C. (A) Auditory encoding customization. A listener can set an encoding channel to map the stock price variable and the scale range for that channel. It is currently set to pitch with a range of 200 hertz to 1,6000 hertz. There is a screenshot of three input forms horizontally arranged. The first form item is a select box labeled Encoding, and pitch is currently selected. The other selectable items are loudness and pan. The second form item is a number input box, labeled From (Stock price: 0), with the current value of 200. The last form item is a number input box, labeled To (Stock price: 230), with the current value of 1,600. (B) Player options. The listener can play and stop the scale description and the audio narrative sonification. These buttons are accessible with a screen reader and keyboard interaction. There is a screenshot showing three buttons for Play scales, Play audio narrative, and Stop audio narrative, aligned from left to right. (C) Audio narrative segments. The listener can also play each segment separately. there are two screenshots connected by ellipsis. Each box consists of text, a line chart, and a play button. The play button in the first box is highlighted and labeled as "C1. play button for a segment." In the first box, the text is 'No. 1. From January first, 2000 to August first, Apple's stock price fluctuates between 21 dollars and 34 dollars.' and the line chart has a wiggly shape. In the second box, the text is 'No. 7. From February first, 2099 to March first, 2010, the stock price steadily increased from 89.31 dollars to 233.02 dollars.'}
    \label{fig:demo:narrative}
\end{figure}

Audio Narrative~\cite{audioNarrative:siu2022} divides a temporal line chart into segments that represent different data patterns, such as increase, decrease, and no change, and offers a sonification and speech description for each segment. 
To show how \emph{Erie} can be used in such applications to generate sonifications, we created an example case that Audio Narrative could create by using \emph{Erie} for sonification and speech generation, as shown in \autoref{fig:demo:narrative}.
We used a \al{`stocks.json'} dataset~\cite{vegadataset} for this replication.
We use two variables, \code{stock price} and \code{date}, from this dataset.

Suppose an Audio Narrative-like application already has a line chart segmented and relevant speech texts generated.
The next task is to create sounds for those segments and speech texts.
Using \emph{Erie}, the application can simply write a sonification spec for each segment as below:
\begin{lstlisting}[name=AN]
{ "description": "...",
  "data": [ /* Segment i data */ ],
  "tone": { "continued": true },
  "encoding": {
    "time": { "field": "date", ... },
    "pitch": { "field": "stock price", ... } } }
\end{lstlisting}
By setting a \code{description}, the application can play the speech for each segment.
Having the above as a template, the application then merge the specs for all the segments as a \code{sequence}:
\begin{lstlisting}[name=AN2]
{ "sequence": [{ /* Segment 1 stream */ }, ... { /* Segment N stream */ }],
  "config": { 
    "forceSequenceScaleConsistency": { "pitch": true },
    "skipScaleSpeech": true
  }}
\end{lstlisting}
The \code{forceSequenceScaleConsistency} in the \code{config} forces the segment streams to use the same \code{pitch} scale. 
As sonifications can benefit from the user's ability to personalize design choices~\cite{sharif2022:sonifier}, we extend this Audio Narrative case by allowing for using a loudness or pan channel to encode a variable and adjusting the scale range of those channels. 
Furthermore, we add an option that separately plays the scale descriptions of a sonification.
\emph{Erie} supports this by using a \code{skipScaleSpeech} option in the \code{config}.

\subsubsection{Chart Reader}\label{sec:demo:reader}

\begin{figure}
    \centering
    \includegraphics[width=\columnwidth]{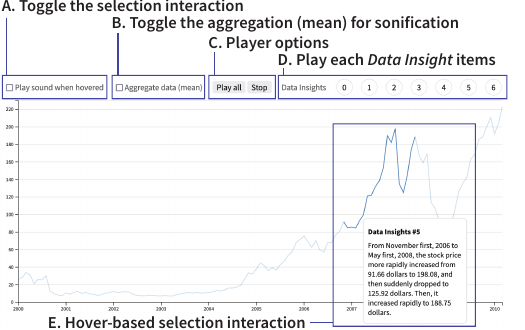}
    \caption{Our replication and extension of Chart Reader~\cite{thompson2023:chartreader} using \emph{Erie}. We further included user options for toggling the hover/selection interaction (A) and aggregation (B).}
    \Description{This figure shows a line chart with several input options on the top and a tooltip on the chart. There are four sections in the input options, from A to D. (A). Toggle the selection interaction. It is a checkbox labeled Play sound when hovered. (B). Toggle the aggregation (mean) for sonification. This is a checkbox labeled Aggregate data (mean). (C). Player options. This consists of two buttons for Play All and Stop. (D). Play each Data Insight item. This consists of seven buttons from 0 to 6.  (E). Hover-based selection interaction. This is the tooltip on the chart. The tooltip content is: 'Data insights No. 5. From November first, 2006 to May first, 2000, the stock price more rapidly increased from 91.66 dollars to 198.08, and then suddenly dropped to 125.92 dollars. Then, it increased rapidly to 188.75 dollars.'}
    \label{fig:demo:reader}
\end{figure}

Given a visualization, Chart Reader~\cite{thompson2023:chartreader} enables hover interaction that reads out values and generates a sonification for the selected data mark(s). 
Furthermore, Chart Reader supports creating several ``data insights'' that allow a sonification author to draft more customized text messages, similar to the chart segments supported by Audio Narrative. 
We replicate this use case by reusing the above Audio Narrative replication, given their underlying structural similarity (segmentation of a chart with descriptive text), as shown in \autoref{fig:demo:reader}.

In this case, the sonification and description text of a chart segment is played whenever the corresponding part in the chart is selected, or the button for the segment is triggered via a mouse or keyboard.
This uses the same segment template spec as Audio Narrative replication, but they are not sequenced.
We set the pitch scale's \code{domain} as the minimum and maximum values of the sonified variable so that the segments can share the same pitch scale even though they are not sequenced in the same specification.
This technique is often used in data visualization cases as well. 
We further include several customization options for toggling the hover interaction and data aggregation.
By reusing the above sequence, we also include an option to play all the `data insight' segments. 

\subsubsection{Nine Rounds a Second}\label{sec:demo:gunman}

The \textit{Nine Rounds a Second} article~\cite{vegas} covers the mass shooting case in Las Vegas in 2017 where the gunman was known to have had a rapid-fire gun. 
This article compares the Las Vegas case with the mass shooting case in Orlando in 2016 and the use of automatic weapons. 
In this article, a dot plot visualizes the shooting count over time to show how fast shots were fired.
To make it even more realistic, the authors of this article included a sonification that mimics actual gun sounds. 

\begin{figure}
    \centering
    \includegraphics[width=\columnwidth]{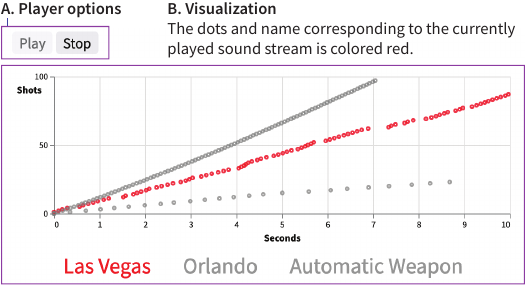}
    \caption{Our replication of the \textit{Nine Rounds a Second} article~\cite{vegas} using \emph{Erie}.}
    \Description{This figure has two sections, A and B. (A). Player options. Two buttons for Play and Stop. (B). Visualization. The dots and name corresponding to the currently played sound stream are colored red. The visualization is a scatterplot for three categories, Las Vegas, Orlando, and Automatic Weapon. Each dot indicates the time and index of a gunshot in each category. The dots for Las Vegas are colored red. They exhibit a long, increasing pattern with short between-dot distances. The dots for Orlando are fewer and more spread with larger between-dot distances. The dots for Automatic Weapon has a highly regular pattern with the same, short between-dot distances. This indicates that the distribution of the Las Vegas dots resembles that of the Automatic Weapon dots.}
    \label{fig:demo:gunman}
\end{figure}

We replicate this news article sonification by mapping the shooting time to a \code{time} channel and the three cases (Las Vegas, Orlando, and automatic weapon) to a \code{repeat} channel, as shown in \autoref{fig:demo:gunman}.
We use an electronic drum's \code{clap} sound that \emph{Erie}'s player supports as a preset because it sounds similar to a gunshot sound.
The original article had an interaction that when the name of a case is selected, it plays only the relevant part. 
To support that, we use the \code{audioQueue.play(i, j)} method, so that the player only plays from the \code{i}-th sub-queue to \code{j}-th sub-queue.
In this case, the first sub-queue is the name of a case, and the last sub-queue is the sonification sound (two sub-queues in total).

\section{Discussion}\label{sec:discussion}
We contribute \emph{Erie}, a declarative grammar for data sonification, with five design goals: independence as a sonification grammar, expressiveness, data-driven syntax, compatibility with audio standards, and extensibility of functionalities.
Below, we briefly discuss remaining technological challenges, and then we motivate future sonification research that could use \emph{Erie}.

\subsection{Technological Hurdles}
While developing \emph{Erie}, we faced two major technical hurdles in using the Web Audio and Speech APIs.
First, there is no standard API that can capture (\ie~generating pure audio files from the source) the sound generated using those APIs.
Instead, users need to use third-party audio capture applications or record sound as it is being played out of the device (which also records room noise and causes distortions due to audio feedback).
Thus, we implemented a workaround Chrome extension\footnote{\url{https://github.com/see-mike-out/erie-chrome-ext}} using Chrome-specific APIs.
Second, speech sounds generated using the Web Speech API cannot overlap which limits \emph{Erie}'s expressiveness, such as the potential to overlay different streams with speeches and tones. 
Thus, related technological extensions to those APIs could help express a more diverse set of audio graphs.

\subsection{Potential Use Cases of \emph{Erie}}
We expect our implementation of \emph{Erie} (compiler, player, and extension APIs) to facilitate various future work on data sonification research and tooling.
In addition to the use cases like sonification for detecting model fit (which might be extended to properties like model convergence), sonification authoring applications, and popular media (\autoref{sec:demo}), future sonification research could use \emph{Erie} to ask questions related to, for example, perceptual intuitiveness and effectiveness of different sonification strategies (\eg~\cite{wang2022:intuitiveness,walker2010:universal}).
Given that sonification design specs expressed in \emph{Erie} can be parameterized as a declarative grammar, sonification researchers could use \emph{Erie} to more systematically generate different stimuli according to their experiment conditions.
Such research will expand understanding around which audio graph formats are best suited for different tasks or auditorily pleasant, providing foundations for building intelligent tools like sonification recommenders.
Furthermore, future sonification tools for data analysis or narrative authoring could use \emph{Erie} as their internal representation to maintain user-specified designs.
To support sonification researchers and developers to test out \emph{Erie}, we provide an online editor for \emph{Erie}\footnote{\url{https://see-mike-out.github.io/erie-editor/}}.

\subsection{Future Work}

\emph{Erie} is our first step of an expressive declarative grammar for data sonification. 
Future work should extend \emph{Erie} to support more dynamic use cases, such as interactivity, streaming data, and different audio environments.

\ipstart{Interactive sonification}
Interactivity is often necessary for data analysis because one static data representation cannot provide a full picture.
While it is possible to use \emph{Erie} in interactive user interfaces with customizability as we demonstrated (\autoref{sec:replication}), \emph{Erie} could better support interactive data sonification with native expressions.
A prerequisite to developing an interactive grammar for data sonification is some understanding of how a sonification listener would trigger a user interaction and receive its feedback using different modalities.
For instance, various approaches to using a keyboard, speech recognition, tabletop screens, or mobile haptic screens for interactive sonification are fruitful topics like personalized sonifications for future research to explore (\eg~\cite{agarwal:sonify,chundury2023:tactualplot}).

\ipstart{Expressing sonifications for streaming data}
Sonification has been used for various real-time streamed data from traditional Geiger counters to audio graphs for physics~\cite{ghosh2010:particle}.
While it is relatively simple for visualization to show existing data points and newly received data points, sonification-based tools may need to build a notion of ``existing'' given the transient characteristic of sound.
For example, a visualization dashboard can express newly received data by adding corresponding visual marks, and the viewers can easily compare them with the existing visual marks. 
However, a sonification monitor may need to play sounds for some past data points, announce the auditory scales, or use notifications for some signals, depending on the task that the listeners want to achieve.
Thus, future work should ask how to indicate and contextualize newly arrived data points, what portion of existing data points should be played again if needed, and how to auditorily imply that a system is waiting on new data.

\ipstart{Supporting different audio environments}
Data sonification can also be useful for other environments like statistical programming and server-side applications. 
For example, \emph{Erie} player for R Studio (a popular integrated development environment for R) could benefit building tools for statistical sonifications like those described above.
As R Studio is backed by Chromium (the same web engine for Chrome and Edge), \emph{Erie}'s web player may need to be extended slightly to support this environment.
To support server-side production of data sonifications, direct generation of raw pulse-code modulation data (digital representation of analog signals)~\cite{pcm} would be useful.

\ipstart{Intelligent authoring tools for data sonification}
As a declarative grammar, \emph{Erie} can make it easier to create data sonifications by allowing developers to declare sonification designs with a few keywords rather than leaving them tedious jobs like inspecting online code and adjusting it to get ad-hoc solutions.
To design effective data sonifications, developers still need to learn relevant knowledge from empirical studies, just as being able to use visualization grammars like D3.js~\cite{battle2022:d3}, Vega-Lite~\cite{satyanarayan:vega-lite2017}, and ggplot2~\cite{wickham:ggplot22010} do not necessarily mean one can easily create effective visualizations.
To support developers in authoring useful sonifications, future work could explore more intelligent approaches like automated design recommenders for different purposes like data analytics, data journalism, and data art.

\subsection{Limitations}
While our primary contribution is the \emph{Erie} grammar, a usable player could make it easier to learn the grammar and apply it to different applications. 
We provide an online player for sonifications backed by \emph{Erie} with baseline functionalities like playing a single queue and showing audio queue tables.
As \emph{Erie} is an open-source project, extensions for more player controls (\eg~playing a single sound) could benefit sonification developers and users with respect to debugging and navigation.
Next, intending \emph{Erie} as a low-level toolkit for sonification developers to use, we prioritized independence from visualization,  expressiveness, and compatibility with audio programming standards.
As \emph{Erie} is not a walk-up-and-use tool, future work could benefit from reflecting on use cases from longer term observations of developer communities.

\section{Conclusion}
\emph{Erie} is a declarative grammar for data sonification design that supports expressing audio channels as data encodings.
\emph{Erie} supports various auditory encoding channels, such as pitch, tapping, and modulation index, and different instruments for sound tones like a simple oscillator, musical instruments, and synths.
Furthermore, we implemented and open-sourced \emph{Erie}'s spec API, compiler, and player for the web audio environment, and they offer extension methods using audio filters and custom encoding channels.
By providing a variety of examples and replicating existing sonification use cases, we demonstrated the expressiveness of \emph{Erie} grammar and the technical feasibility of our implementations.
We expect \emph{Erie} to support various data sonification research and produce further understanding in auditory perception of data, which will in turn help extend \emph{Erie}'s capabilities.



\bibliographystyle{ACM-Reference-Format}
\bibliography{bibliography}


\appendix
\section{Technical Details of \emph{Erie}}\label{sec:appendix}
This appendix details the \emph{Erie} grammar.

\subsection{Customizing a Tone}\label{sec:appendix:tone}
The \al{tone} of a single sonification design is defined in terms of instrument \al{type}, whether the sound is \al{continued}, and a set of audio \al{filter}s.
To use custom instruments to express diverse sonification designs, a developer can define new instruments using \al{synth} (for FM and AM synthesizer), \al{wave} (directly defining a wave function), and \al{sampling} (importing external audio files) objects in a top-level spec.
The developer can apply custom instruments to the \al{tone} \al{type} and a \al{timbre} encoding channel by referencing their names. 

A dataset typically exists as a set of data points; even if it represents a continuous distribution, its digital format is a set of approximated data points. 
Thus, a data representation should be able to capture the continuity or discreteness between data points (\eg~line chart vs. scatterplot). 
In the walkthrough, for example, we used a \al{discrete} ($\textit{continued} = false$) tone to indicate that each sound represents a discrete bin.
On the other hand, a developer could use a \al{continuous} ($\textit{continued} = true$) tone for a sonification of a continuous distribution.
A sound is \al{discrete} if it is momentarily paused and resumed as auditory values change (\eg~a sound ``beep Beep BEEP'' with varying loudness).
A sound is \al{continuous} if it is not paused as its auditory values change (\eg~a sweeping sound ``bee$^{C3}$-ee$^{C\#3}$-eep$^{D3}$'' with increasing pitch).

When more artistic sound effects (\eg~dynamic compression, distortion) are needed, a developer can apply them using the \al{filter} property of a \al{tone}.
A \al{filter} object is an ordered list of filter names, and each filter is applied after the previous filter, reflecting how audio filters are commonly applied to electric sounds.
\emph{Erie} considers the properties of an audio filter (\eg~level of compression) as encoding channels so that a developer can configure audio filters both statically and dynamically (mapped to data variables).
Our implementation offers several preset filters (\eg~dynamic compressor) and APIs for audio experts to define and use custom filters.

\subsection{Encoding}\label{sec:appendix:encoding}

Below, we detail how to indicate specific properties for different encoding channels and auxiliary or shortcut properties for diverse sonification designs.

\subsubsection{Expressing time as an encoding}\label{sec:appendix:time}
Time is to sonification as position is to visualization.
An audio graph arranges its auditory values along a time axis.
\emph{Erie} expresses time as encoding to enable data-driven time control.
For example, the start time of each sound can be mapped to a certain data variable. 

The time axis of a sonification encodes data either in terms of the start and end times of a sound (\al{time} and \al{time2}) or the start time and duration of a sound (\al{time} and \al{duration}).
On the one hand, two data variables sharing the same unit (\eg~monthly lowest and highest temperature) can be mapped to start and end times. 
On the other hand, two data variables with different units (\eg~country names and CO2 emissions) can be mapped to start time and duration. 
\emph{Erie} supports expressing when a sound starts and ends (\al{time} + \al{time2}) or when and how long it is played (\al{time} + \al{duration}).

The length of a sonification is also the \al{range} of its time channel.
Thus, \emph{Erie} provides another shortcut, \al{length}, for the \al{range} of time scale (\ie~$[0, \textit{length}]$). 
When there is no need to encode end time or duration, \al{time} channel can have \al{band} to set the duration of each sound uniformly (for discrete tones).

\emph{Erie} makes a distinction between \textit{when a sound starts} (the value of the \al{time} channel) and \textit{how a sound is timed} in relation to other sounds (\al{timing}). 
For example, a developer wants a sound to be played after the previous sound (\al{relative}), to start on an exact time (\al{absolute}), or to start with the other sounds (\al{simultaneous}). 
To control how a \al{time} channel assigns times, the developer can use the \al{timing} property of the \al{time} channel's scale.
The above extensions to the \textit{time} channel's scale is formalized as:
\begin{align*}
\textit{scale}_{\textit{time}} := \{ \ldots, \textit{timing}, \textit{length}, \textit{band} \}
\end{align*}

These time-related channels and the \al{timing} option produce the following high-level combinations:

\istart{Case 1} $\textit{time}(\textit{field}=x, \textit{scale.band}=\textit{n})$.
A \al{time} channel with a fixed \al{scale.band} value defines sounds with a fixed duration ($n$) and start times varied by the encoded data variable ($x$).
If \al{scale.timing} is \al{simultaneous}, then all of the sounds are played at the same time.

\istart{Case 2} $\textit{time}(\textit{field}=x) + \textit{duration}(\textit{field}=y)$.
Using both \al{time} and \al{duration} channels defines sounds with varying durations and start times.

\istart{Case 3} $\textit{time}(\textit{field}=x, \textit{scale.timing}=absolute) + \textit{time2}(\textit{field}=y)$.
A \al{time} channel with \al{absolute} \al{timing} and a \al{time2} channel specify sounds with varying start and end times.
Note that the two fields mapped to the \al{time} and \al{time2} channel must be defined on the same data unit, such as bin start and endpoints. 

\subsubsection{Channel-specific properties}\label{sec:appendix:channel}
Specific auditory encoding channels may have different physical constraints that need channel-specific properties in addition to the above definition.
\emph{Erie} considers such physical constraints in defining encodings for canonical auditory channels.
For example, \al{pitch} can have raw pitch frequency values or have them rounded to musical notes.
To enable this rounding, a developer can set \al{round-to-note} to \al{true} for the \al{pitch} channel.

\subsubsection{Providing auditory reference elements}\label{sec:appendix:reference}

\ipstart{Tick for time channel}
A longer sonification may need to provide a sense of the progression of time as Cartesian plots have axis ticks and grids.
To do so, a developer could use a \al{tick} sound that repeats every certain time interval.
The developer can define a \al{tick} directly in the \al{time} channel or refer to a tick definition in the top-level \al{tick}.

\ipstart{Scale description markup}
As we use legends for data visualizations, it is important to provide the overview of the scales used in a sonification~\cite{wang2022:intuitiveness}.
The \al{description} of a scale can be skipped, defined as a default audio legend set by a compiler, or customized.
To customize a scale description, \emph{Erie} employs a markup expression that can express literal texts, audio legends (\al{<sound>}), a list of items (\al{<list>}), and reserved keywords, such as \al{<domain.min>} (for the minimum domain value) and \al{<field>} (for the data field's name).
A developer can also pass a number or date-time \al{format} in the channel definition.

\subsubsection{Inline Transform}\label{sec:appendix:inline}
Inspired by Vega-Lite~\cite{satyanarayan:vega-lite2017}, it is possible to provide an inline data transform: \al{aggregate} or \al{bin}.
This is a shortcut for defining a corresponding \al{transform} item and use the resulting variables in the channel's \al{field}. 
For example, the separately defined transforms in the walkthrough can be rewritten as:
\begin{align*}
\textit{time} & =\{\textit{field}=\eqValue{miles-per-gallon}, \textit{bin}=\eqValue{true}, \cdots\} \\
\textit{pitch} & =\{\textit{aggregate}=\eqValue{count}, \cdots\}
\end{align*}


\end{document}